\documentclass[12pt,preprint]{aastex}

\newcommand{\swift}{{\it Swift}}

\slugcomment{ApJ in press}

\shorttitle{Canonical GRB afterglows with \swift}
\shortauthors{}

\begin{document}

\title{Evidence for a Canonical GRB Afterglow
Light Curve in the {\it \bfseries Swift} XRT Data
}

\author{
J.~A.~Nousek\altaffilmark{1},
C.~Kouveliotou\altaffilmark{2},
D.~Grupe\altaffilmark{1},
K.~L.~Page\altaffilmark{3},
J.~Granot\altaffilmark{4},
E.~Ramirez-Ruiz\altaffilmark{5,6}, 
S.~K.~Patel\altaffilmark{7,2},
D.~N.~Burrows\altaffilmark{1},
V.~Mangano\altaffilmark{8},
S.~Barthelmy\altaffilmark{12},
A.~P.~Beardmore\altaffilmark{3},
S.~Campana\altaffilmark{9},
M.~Capalbi\altaffilmark{10},
G.~Chincarini\altaffilmark{9,11},
G.~Cusumano\altaffilmark{8},
A.~D.~Falcone\altaffilmark{1},
N.~Gehrels\altaffilmark{12},
P.~Giommi\altaffilmark{10},
M.~R.~Goad\altaffilmark{3},
O.~Godet\altaffilmark{3},
C.~P.~Hurkett\altaffilmark{3},
J.~A.~Kennea\altaffilmark{1},
A.~Moretti\altaffilmark{9},
P.~T.~O'Brien\altaffilmark{3},
J.~P.~Osborne\altaffilmark{3},
P.~Romano\altaffilmark{9},
G.~Tagliaferri\altaffilmark{9}, 
A.~A.~Wells\altaffilmark{3}
}

\altaffiltext{1}{Department of Astronomy \& Astrophysics, Pennsylvania State
	University, University Park, PA 16802, USA}
\altaffiltext{2}{NASA/Marshall Space Flight Center, National Space
  Science Technology Center, XD-12, 320 Sparkman Drive, Huntsville, AL
  35805, USA.}
\altaffiltext{3}{Department of Physics and Astronomy, University of Leicester,
	Leicester, UK}
\altaffiltext{4}{Kavli Institute for Particle Astrophysics and Cosmology, Stanford University, P.O. Box 20450, MS 29, Stanford, CA 94309.}
\altaffiltext{5}{Institute for Advanced Study, Einstein Drive, Princeton, NJ 08540, USA.}
\altaffiltext{6}{Chandra Fellow.}
\altaffiltext{7}{IPA with NASA/MSFC through Universities Space Research Association}
\altaffiltext{8}{INAF- Istituto di Fisica Spazialee Fisica Cosmica sezione di Palermo, Palermo, Italy}
\altaffiltext{9}{INAF -- Osservatorio Astronomico di Brera, Merate, Italy}
\altaffiltext{10}{ASI Science Data Center, Frascati, Italy}
\altaffiltext{11}{Universit\`a degli studi di Milano-Bicocca,
  Dipartimento di Fisica, Milan, Italy}
\altaffiltext{12}{NASA/Goddard Space Flight Center, Greenbelt, MD}

\begin{abstract}

We present new observations of the early X-ray afterglows of the first
27 gamma-ray bursts (GRBs) well-observed by the {\it Swift} X-ray
Telescope (XRT). The early X-ray afterglows show a canonical behavior,
where the light curve broadly consists of three distinct power law
segments: (i) an initial very steep decay ($\propto t^{-\alpha}$ with
$3\lesssim\alpha_1\lesssim 5$) , followed by (ii) a very shallow decay
($0.5\lesssim\alpha_2\lesssim 1.0$), and finally (iii) a somewhat
steeper decay ($1\lesssim\alpha_3\lesssim 1.5$). These power law
segments are separated by two corresponding break
times, $t_{\rm break,1}\lesssim 500\;$s and
$10^3\;{\rm s}\lesssim t_{\rm break,2}\lesssim 10^{4}\;$s. On top
of this canonical behavior, many events
have superimposed X-ray flares, which are most likely caused by
internal shocks due to long lasting sporadic activity of the central
engine, up to several hours after the GRB. We find that the initial
steep decay is consistent with it being the tail of the prompt
emission, from photons that are radiated at large angles relative to
our line of sight.  The first break in the light curve ($t_{\rm
break,1}$) takes place when the forward shock emission becomes
dominant, with the intermediate shallow flux decay ($\alpha_2$) likely
caused by the continuous energy injection into the external
shock. When this energy injection stops, a second break is then
observed in the light curve ($t_{\rm break,2}$).  This energy
injection increases the energy of the afterglow shock by at least a
factor of $f\gtrsim 4$, and augments the already severe requirements
for the efficiency of the prompt gamma-ray emission.

\end{abstract}


\section{Introduction\label{introduction}}

Gamma-ray bursts (GRBs) are believed to arise from the sudden release
of a vast amount of energy ($\sim 10^{51}\;$erg for long GRBs and
probably somewhat less energy for short GRBs), into a very small
region (of size $\lesssim 10^2\;$km), over a short time ($\sim
10-10^2\;$s for long GRBs and $\lesssim 2\;$s for short GRBs).  The
prompt $\gamma$-ray emission is attributed to internal shocks within
the outflow \citep{RM94,SP97} that are caused by variability in its
Lorentz factor, $\Gamma$.  The highly non-thermal gamma-ray spectrum
requires $\Gamma\gtrsim 10^2$ to avoid the {\it compactness problem}
\citep[see][and references therein]{Piran99}. When the relativistic
ejecta sweep up a sufficient amount of external medium, they are
decelerated by a (typically mildly relativistic) reverse shock; at the
same time a highly relativistic forward shock is driven into the
ambient medium. This forward shock produces the long-lasting afterglow
emission, while the short lived reverse shock produces a prompt
optical emission ({\it optical flash}); the latter peaks on a
time scale of tens of seconds \citep{Akerlof99,SP99,FRW99,SR02} and
dominates the optical emission up to about ten minutes.

GRBs naturally divide into two classes according to their duration and
the hardness of their prompt gamma-ray emission: short/hard bursts
($<2\;$s) and long/soft bursts ($>2\;$s) \citep{Kouveliotou93}. Until
very recently, afterglow emission was detected only for long GRBs,
leading to significant progress in their understanding. To date over
50 spectroscopic redshifts have been determined (typically
$0.5\lesssim z\lesssim 3$) and an association of some long GRBs with a
contemporaneous Type Ic supernova \citep{Galama98,Stanek03,Hjorth03,Malesani04} 
has been established, which implies a massive star as the long GRB progenitor
and thus, supports the collapsar model \citep{Woosley93}. Until the
launch of \swift\ \citep{Gehrels04}, the progenitors of short GRBs 
remained largely a mystery. 
The leading short GRB model featuring the merger of a compact
binary, usually NS-NS \citep[e.g.][]{Eichler89} or NS-BH
\citep[e.g.][]{Pac91}, was given strong support recently with the
detection of afterglow emission from three short GRBs: 050509B,
050709, and 050724
\citep{Bloom05,Gehrels05,Hjorth05,Lee05,Berger05,Covino05,Barthelmy05}.

Before \swift, X-ray afterglow emission was detected, in most cases,
only several hours after the burst, by which time the flux typically
showed a smooth single power law decay $\sim t^{-1}$. In contrast, the
optical afterglow light curve often showed an achromatic steepening
to $\sim t^{-2}$, attributed to a narrow jet whose edges become
visible as it decelerates and widens \citep{Rhoads99,SPH99}. However,
the {\it early} afterglow evolution -- the first few hours, which can
probe important questions such as the density profile of the external
medium and the early radiative energy losses from the external shock --
remained largely unexplored.  \swift\ is designed to, among other
science, probe exactly this unknown observational time window from
$\sim 10^2\;$s to $\sim 10^4\;$s after burst onset. Here we report for
the first time cumulative early X-ray afterglow properties of the
first 27 {\it long} GRBs well-observed by the \swift/XRT.\footnote{The
XRT also observed GRBs 050117, 050306, 050416B and 050528.  The first 
was observed while the XRT was in a high particle background; the middle
two were observed days after the burst due to observing constraints;
the last was observed while XRT was in an engineering mode.
\citet{Hill05} have reduced the data for GRB050117, and find a similar
light curve to the canonical behavior described here.
GRB050509B is a
{\it short} burst, and not included for that reason \citep{Gehrels05}.}
In \S\ref{data}
we describe our data analysis method. Our observational results are
presented in \S \ref{obs}. In \S \ref{theory} we discuss the
theoretical interpretation and implications of our findings, and our
conclusions are summarized in \S \ref{conc}.

\section{Data Analysis}
\label{data}

We have analyzed XRT data of the first 27 \swift\ GRB afterglows
covering the time interval between December 2004 - June 2005. Data for
each burst were obtained from the \swift\ Quick Look
site\footnote{http://heasarc.gsfc.nasa.gov/cgi-bin/sdc/ql?}
and processed consistently with version 2.0 of the \swift\ software 
(release date 2005-04-05). In all cases we used XSELECT to extract source and
background counts from the cleaned event lists ($0.3-10\;$keV), using
grades 0-12 for Photon Counting (PC) mode data, 0-2 for Windowed
Timing (WT) and 0-5 for Photo-Diode (PD). We used the European
Southern Observatory (ESO) Munich Imaging Data Analysis System (MIDAS,
version 04SEP), to create the X-ray afterglow light curves for each
event. The data were binned dynamically to have a certain number of
photons per bin. For very bright bursts and at early times after a
burst trigger, the binning was set to 500 photons per bin, while at
very late times, or for very faint bursts, the binning was set to 10
counts per bin. On average, light curves were created with 50
counts per bin. All light curves were background subtracted. The
exposure times per bin were calculated on the basis of the Good Time
Interval (GTI) file. These light curves were then compared to ones
derived independently with the {\sc ftool} {\bf flx2xsp}. Each time
bin in the latter was selected for high signal to noise ratio, after
background subtraction; we required at least 20 counts per bin in
order to facilitate $\chi^2$ fitting. The data sets derived using
these two independent methods were found to agree very well. Finally,
in both methods, we took into account the mode switching during the
\swift/XRT observation, which can distort the real count rate during
an event.

Several of the GRBs included in this paper were observed while \swift\
was still in its calibration phase, before the automatic
mode-switching for the XRT was fully enabled. Some of the data
obtained in PC mode suffered, therefore, from pile-up, which had to be
corrected before the light-curves and spectra were fully
analyzed. To account for source pileup (significant above 0.5 counts/s
in Photon Counting mode), annular regions were used to extract the
source spectra and light-curves. To determine the level of pile-up,
the inner radius of the annulus was gradually increased until the
spectral shape no longer changed (pile-up leads to the hardening of
photon indices). Background spectra and light-curves were then
produced from large `source-free' regions, offset from the GRB, and
the background counts were scaled to the same size region as used for
the source.

The {\sc ftool} {\bf xrtmkarf} was used to generate ancillary response
function (ARF) files. Where an annular region had been required, {\bf
xrtmkarf} was run twice, with and without the Point Spread Function
(PSF) correction. Fitting the spectra with both ARFs leads to
different normalizations, the ratio of which gives the pile-up
correction factor. The most recent (version 7) response matrices
(RMFs) were used in the spectral analysis. The light-curves were
extracted for each individual orbit of data, correcting for pile-up
when annuli were used. At later times, or when no pile-up was
apparent, circles of radius 20-30 pixels (1 pixel =
2.36$^{\prime\prime}$) were used.

The XSPEC (version 11.3.2) readable light-curves produced by {\bf flx2xsp}
were modeled in XSPEC with a combination of single and
broken power-laws to determine the decay slopes and break times. The
time of the burst onset was taken from the msbal.fits TDRSS file,
which normally corresponds to the time when the BAT instrument recognized
the burst through an image trigger,
except for the case of GRB~$050319$, where the event started while
\swift\ was slewing to a different target (although triggers are
disabled during slews, the BAT triggered on a later peak in the
light-curve of GRB~$050319$). To determine an energy conversion factor
(ECF) from count-rate to fluxes, a simple absorbed \citep[Galactic
$N_{\rm H}$, determined from][together with any required
excess]{dickey90} power-law was fitted to the XRT spectra ($0.3-10$
keV). The ECFs were then determined for unabsorbed fluxes. If no
significant spectral changes were observed, only one ECF was applied
per light curve.

\section{Observational Results}
\label{obs}

Until July 2005, only 10 \swift\ GRBs had measured redshifts. Figure
\ref{L_X(t)} exhibits the evolution of the X-ray luminosities of these
10 \swift\ events together with the longest monitored GRBs in the last
8 years \citep[see also][]{kouv04}. The \swift\ light curves fill in
the earlier gap and complete the trend observed in the past
\citep{kouv04} in a spectacular way. Fig.~\ref{F_X(t)} shows the
evolution of the X-ray flux for the 17 \swift\ GRBs without known
redshifts. Four of these events show X-ray flares early on ({\it lower
panel} of Fig.~\ref{F_X(t)}).

\begin{figure}[h]
\setlength{\unitlength}{1.0in}
\begin{picture}(6,7.5)
\put (0.0,0.0){\centerline{\includegraphics[width=0.6\textwidth]{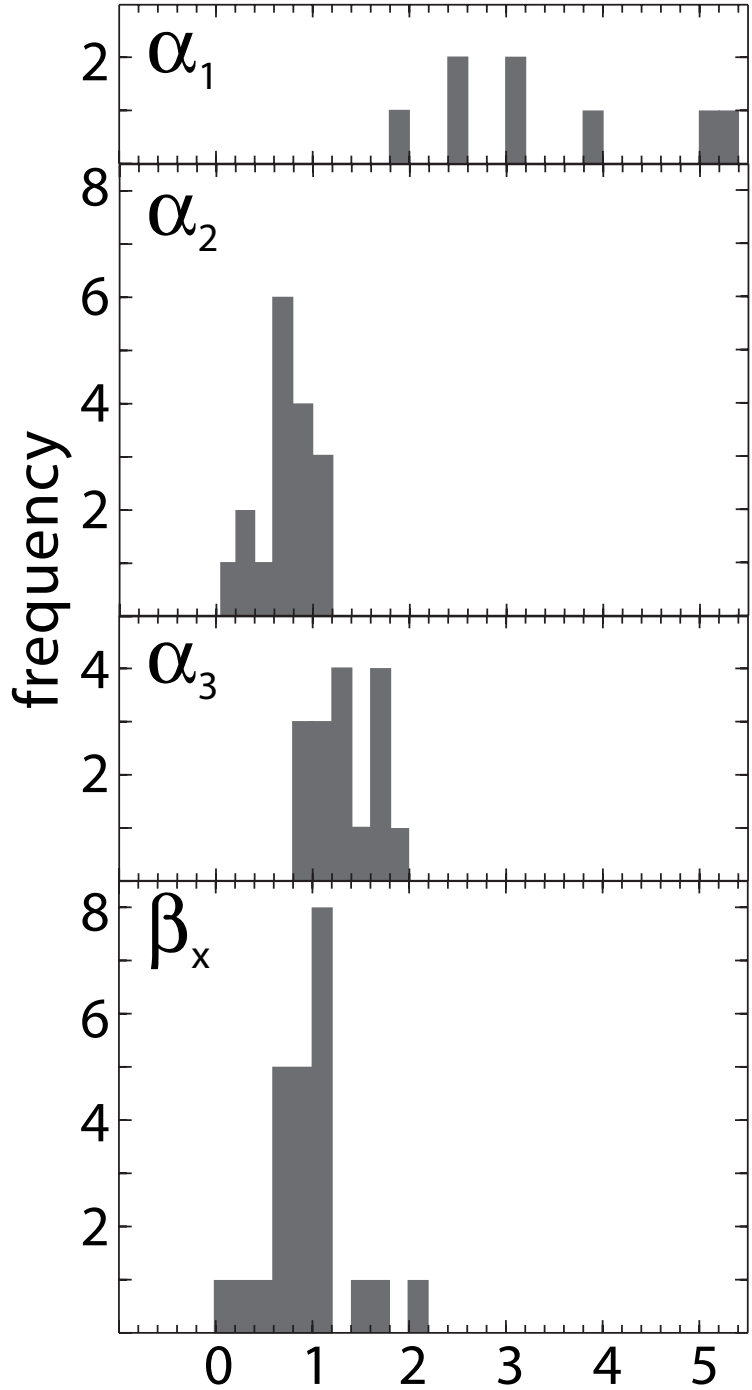}}}
\end{picture}
\caption{\label{alpha123}Histogram of the spectral index $\beta_{\rm
  x}$ and the temporal indices $\alpha_{1}$, $\alpha_2$ and
  $\alpha_3$, for the GRBs in Table 1. Note that only $\beta_{1,\rm
  x}$ is plotted here for the events with evolving spectral
  properties. The x-scale range is the same for all indices.}
\end{figure}

Combining Figs. \ref{L_X(t)} and \ref{F_X(t)} we see that a general
trend starts to emerge that may become the standard to describe each
GRB X-ray afterglow light curve.  Starting at the earliest XRT
observations (approximately $10^2\;$s after the prompt gamma-rays),
the X-ray flux $F_\nu$ follows a canonical behavior comprising three
power law segments where $F_\nu\propto\nu^{-\beta}t^{-\alpha}$ (see
also Fig. \ref{diagram}): an initial steep decay slope ($\alpha_1$),
which (at $t_{\rm break,1}$) changes into a very flat decay
($\alpha_2$), that in turn (at $t_{\rm break,2}$) transitions to a
slightly steeper slope ($\alpha_3$).  Table 1 lists the temporal and
spectral parameters for all 27 events, as well as the break times
($t_{\rm break,1}$ and $t_{\rm break,2}$), the BAT trigger times, and
the onset of the XRT data after trigger. The spectrum remained
constant throughout the breaks (within our available statistics) in
all cases except two (GRBs 050315 and 050319) where the spectrum
hardened (i.e. $\beta$ decreased) across the first break (at $t_{\rm
break,1}$). Fig.~\ref{alpha123} shows the distribution of all temporal
indices ($\alpha_1$, $\alpha_2$ and $\alpha_3$) together with the
spectral index ($\beta_{\rm X}$) for the GRBs in Table 1. 

Table 1 shows that in about half ($40.7\%$) of the cases we were able to slew to
the initial BAT location with XRT only several thousands of seconds
after the GRB trigger; in the majority of these cases we detect
one or even no temporal break in the X-ray afterglow light curve. In
the latter cases (without a break) we have defined as $\alpha_3$ the
slope that prevails beyond $2\times 10^4\;$s in a light curve. It should be
noted here that the values of $\alpha_3$ are consistent with those
seen in previous missions since they typically started observations
hours after the burst.

\begin{figure}[h]
\setlength{\unitlength}{1.0in}
\begin{picture}(6,7)
\put (-0.02,0.0) {\centerline{\includegraphics[width=0.85\textwidth]{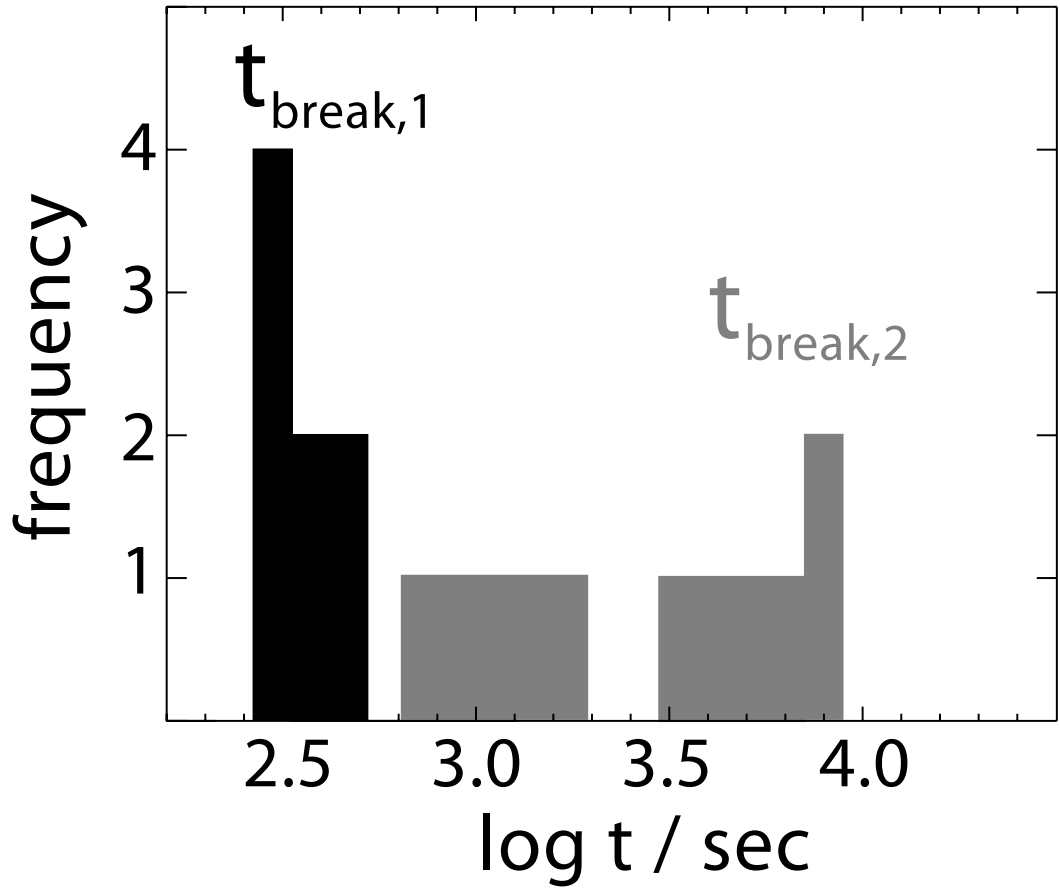}}}
\end{picture}
\caption{\label{tbreaks}Histogram of $t_{\rm break,1}$
  and $t_{\rm break,2}$.}
\end{figure}

Whenever we found early ($<500\;$s) breaks in the light curves of GRBs
with established redshifts, we converted them to the GRB cosmological
rest frame (below, but not in Table 1). We have three such cases, GRBs
050126, 050315, and 050319, with rest frame breaks at 185, 136, and 87
seconds, respectively. This sample, together with GRBs 050219A ($\leq
332\;$s) and 050422 ($\leq 272\;$s), strongly point to an early X-ray
afterglow light curve break, $t_{\rm break,1}\lesssim 300\;$s in the
cosmological rest frame (or $t_{\rm break,1}\lesssim 500\;$s in the
observer frame). The distribution of $t_{\rm break,1}$ and $t_{\rm
break,2}$ (without correcting for cosmological time dilation, since
the redshift is not known for most of the GRBs in our sample) is shown
in Figure \ref{tbreaks}.

We proceed to calculate the observed X-ray flux ($2-10\;$keV) at
$1\;$hr and $10\;$hr after the GRB trigger; whenever there was no
direct measurement of the flux, we have used the temporal parameters
of Table 1 to extrapolate to these times, using as a starting point
the Spacecraft Clock trigger times (Table 1). We have used luminosity
distances, $d_L$ for $(\Omega_M,\Omega_\Lambda,h)=(0.3,0.7,0.71)$, and
spectral parameters in order to calculate the isotropic equivalent
X-ray luminosities at $1\;$hr ($L_{\rm X,1}$) and $10\;$hr ($L_{\rm
X,10}$) and isotropic equivalent energy output in gamma-rays
($E_{\rm\gamma,iso}$). For those GRBs with reshifts the luminosity is:
\begin{equation}\label{L_X}
L_{\rm X}(t) \equiv \int_{\nu_1}^{\nu_2}d\nu\,L_\nu(t) = 
\frac{4\pi d_L^2}{(1+z)}\int_{\nu_1}^{\nu_2}d\nu\,F_{\nu/(1+z)}[(1+z)t] 
= 4\pi d_L^2\int_{\nu_1/(1+z)}^{\nu_2/(1+z)}d\nu\,F_\nu[(1+z)t]\ ,
\end{equation}
where $L_\nu(t)$ is the spectral luminosity at the cosmological rest
frame of the source (i.e. both $\nu$ and $t$ in $L_\nu(t)$ are
measured in that frame), while $F_\nu(t)$ is measured in the
observer's frame. When assuming $F_\nu \propto
\nu^{-\beta}t^{-\alpha}$, Eq. \ref{L_X} simplifies to $L_{\rm X}(t) =
4\pi d_L^2(1+z)^{\beta-\alpha-1}F_{\rm X}(t)$, where $F_{\rm X}(t) =
\int_{\nu_1}^{\nu_2}d\nu\,F_\nu(t)$.

The corresponding $L_{\rm x,1}$ and $L_{\rm x,10}$ ($2-10$ keV) were then
calculated using the relevant spectral and temporal indices listed in
Table 1. These values are listed in Table 2 together with the
K-corrected values of $E_{\rm\gamma,iso}$ for each GRB. The latter has
been recalculated within two energy bands, the narrower of which
($100-500\;$keV) overlaps in all GRBs. The wider band,
($20-2000\;$keV), is an upper limit and assumes no spectral changes
from a single power law fit in the GRB prompt emission.

\begin{figure}[ht]
\setlength{\unitlength}{1.0in}
\begin{picture}(6,7)
\put (-0.02,0.0) {\centerline{\includegraphics[width=0.95\textwidth]{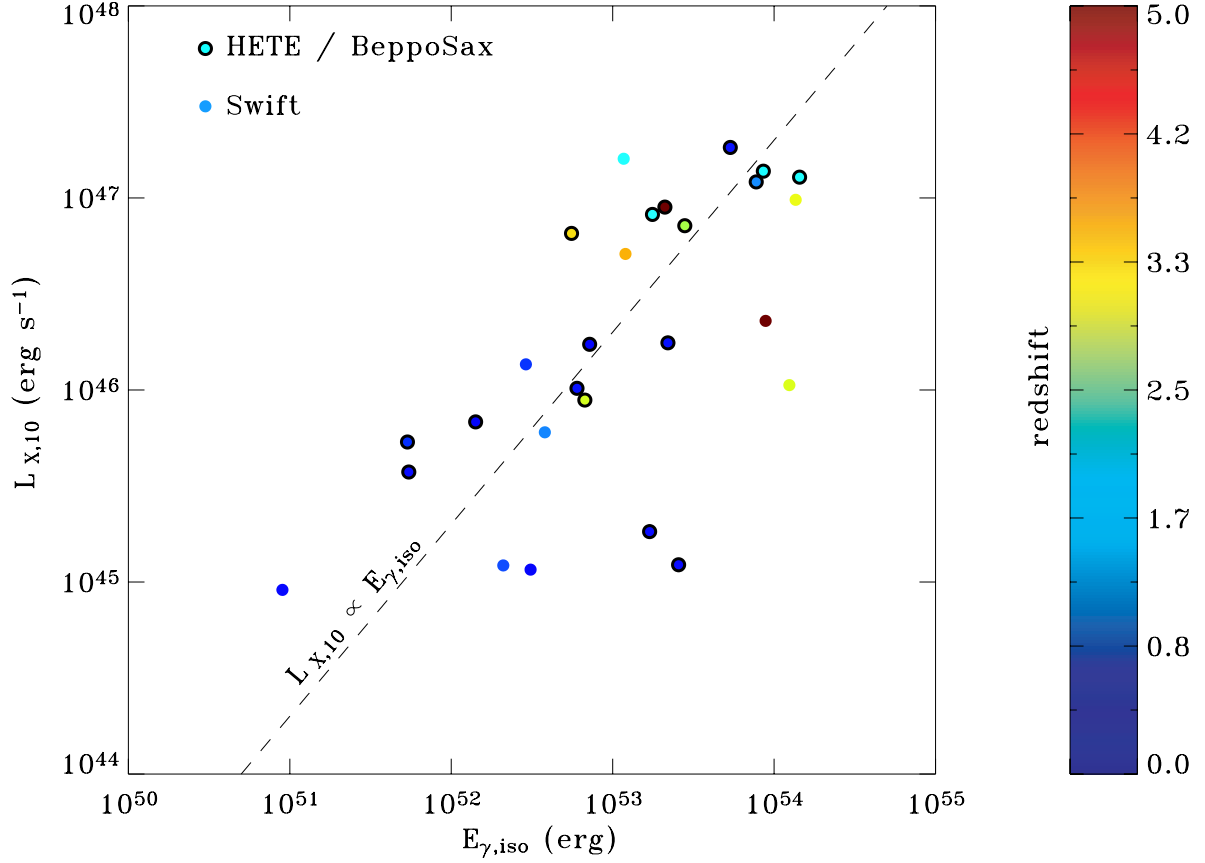}}}
\end{picture}
\caption{\label{L_x-E_iso}Distribution of $L_{\rm X,10}$ ($2-10$ keV) versus
  $E_{\rm\gamma,iso}(20-2000\,{\rm keV})$ for all {\it Swift} GRBs
  with established redshifts (from Table 1) plotted together with
  selected earlier events observed with {\it HETE-II} and {\it
  BeppoSAX} \citep{Berger03,Bloom03}.}
\end{figure}

Figure \ref{L_x-E_iso} shows $L_{\rm X,10}$ versus $E_{\rm\gamma,iso}$
($20-2000\;$keV) for the 10 \swift\ GRBs with established redshifts,
as well as 17 {\it HETE-II} and {\it BeppoSAX} GRBs ($20-2000\;$keV),
for comparison.  The distribution of Swift events is compatible with 
that of earlier events measured with HETE-II and BeppoSAX; the combined 
sample is consistent with an apparent positive, roughly linear, 
correlation between $L_{\rm X,10}$ and $E_{\rm\gamma,iso}$. 
(We have calculated the
linear correlation and Spearman rank order correlation and find that random
chance would have produced the observed values only 15\% and 7\% of the time,
respectively.)
We assume that $E_\gamma \gtrsim E_{kin,afterglow}$ for most GRBs 
\citep[see][for support of this idea]{FriWax01}.
If, also, energy injection occurs in most GRBs,
then the observation that the ratio of $E_{\gamma,iso}$ to $L_{X,10}$
in our sample is similar to that in pre-Swift GRBs implies that 
$E_\gamma \gtrsim E_{kin,afterglow}$ for our sample as well.
This suggests a high efficiency for the prompt gamma-ray emission,
which is roughly constant (albeit with large scatter).
Figure \ref{L_x-E_iso} also contains a color coding for the
redshift of the different events; we note here an apparent positive
correlation between $E_{\rm\gamma,iso}$ (or $L_{\rm X,10}$) and the
redshift, $z$. This is likely due to observational selection effects,
since, at least on average, intrinsically dimmer (brighter) events can
be detected, their X-ray luminosity measured, and their redshift
determined, out to a smaller (larger) redshift.

\section{Theoretical Interpretation of the X-ray afterglow light curve}
\label{theory}

\subsection{The Early Rapid Flux Decay ($\alpha_1$)}
\label{alpha1}
 
The most natural explanation for the early rapid flux decay with
$3\lesssim\alpha_1\lesssim 5$ is emission from {\it naked GRBs}
\citep{KP00}, i.e. prompt GRB emission (that is usually attributed to
internal shocks) from large angles ($\theta>\Gamma^{-1}$) relative to
our line of sight that reaches us at late times ($\Delta t\approx
R\theta^2/2c$), resulting in a steep flux decay with
$\alpha_1=2+\beta_1$. This relation is more or less satisfied in most
(though not all) cases for which $\alpha_1$ could be determined (see
Figure \ref{alpha1beta1}).  Note also that \citet{Barthelmy05} 
have studied the BAT and XRT spectral parameters for GRBs 050315 and 050319, and
conclude that the same spectrum is consistent with both the BAT burst
data and the early XRT afterglow data.

A somewhat steeper power law decay ($\alpha_1>2+\beta_1$) can be
obtained within a few $T_{\rm GRB}$, where $T_{\rm GRB}$ is the
duration of the prompt emission, for the following reason. The
temporal decay index of $\alpha_1=2+\beta_1$ applies separately to
each spike in the prompt light curve, as it corresponds to a collision
between two sub-shells in the internal shocks models, where the
emission from that collision decays as
$F_\nu\propto(t-t_0)^{-\alpha_1}$. Here the reference time, $t_0$, of
the power-law decay corresponds to the onset of that particular spike
(i.e. the time at which the outer of the two sub-shells was ejected
from the source). Since all power law fits to the light curve take the
GRB trigger (which corresponds to the onset of the first spike) as the
reference time, this would cause a seemingly steeper power law decay
index for later spikes. The decay of the last spike, for which
$t_0\approx T_{\rm GRB}$, will approach a power-law decay in $t$ for
$t/T_{\rm GRB}\gtrsim{\rm a\ few}$. This would lead to a decrease in
$\alpha_1$ with time until it approaches $2+\beta_1$ at $t/T_{\rm
GRB}\gtrsim{\rm a\ few}$. Thus, if $t_{\rm break,1}/T_{\rm
GRB}\lesssim\;{\rm a\ few}$, the asymptotic values of
$\alpha_1=2+\beta_1$ might not be reached. On the other hand, a
shallower temporal decay index, $\alpha_1<2+\beta_1$, is hard to
achieve in this scenario, and might require a different physical
origin. In practice, however, our mostly sparse coverage of the XRT
light curves at $t<t_{\rm break,1}$ might lead to an underestimation
of $\alpha_1$, since the value that is derived from the fit to the
data may not represent its asymptotic value at $t\ll t_{\rm break,1}$.

Assuming that the {\it naked GRB} interpretation is correct, we expect
$F_\nu(t<t_{\rm break,1})$, or its extrapolation back in time, to
smoothly join with the prompt GRB emission as the emission immediately
after the burst would be dominated by the tail of the last spike. At
later times, $t\gtrsim 2T_{\rm GRB}$, the light curve would have
contributions from the tails of all spikes with a relative weight
similar to that of the spikes themselves. 
The bursts for which we have 
direct temporal overlap between the BAT and XRT are consistent with a smooth
connection in flux \citep[e.g.,][]{Vaughan05}.  
\citet{O'Brien05} have considered
40 Swift bursts with prompt XRT observations and conclude that the
BAT and XRT join smoothly, although in a few cases 
\citep[such as GRB050219a;][]{Tagliaferri05}
the smooth connection is confused by the presence of an early X-ray flare in the
light curve.

An interesting alternative model for the initial fast decay, that
might apply at least in some cases, is reverse shock emission from
large angles relative to our line of sight \citep{Kobayashi05}. This
emission might be either synchrotron or synchrotron self-Compton
(SSC). The latter could suppress the flux in the optical relative to
that in the X-rays \citep{Kobayashi05}, thus supporting the strict
upper limits on the early optical flux that exist for some of the GRBs
in our sample. This interpretation would require, however, a large
Compton y-parameter, and in turn a very low magnetization of the GRB
outflow. The synchrotron component of the reverse shock emission could
dominate in the X-ray range. This is theoretically possible despite
the fact that the $F_\nu$ spectrum peaks around the optical or IR,
since the $\nu F_\nu$ spectrum peaks closer to the X-ray range and is
fairly flat above its peak, so that a good fraction of the total
emitted energy can fall within the X-ray range.

Finally, several other models can also be considered to explain this
part of the X-ray light curve \citep{Tagliaferri05}. For example,
emission from the hot cocoon in the context of the collapsar model
\citep{MR01,R-RCR02} might produce a sufficiently steep flux decay,
but would naturally produce a quasi-thermal spectrum which does not
agree with the observed power-law spectrum.  Photospheric emission as
the ejecta becomes optically thin \citep{RM05,RR05} is also possible
as it may be able to produce significant deviations from a thermal
spectrum although it is unclear how this emission would last longer
than the prompt gamma-ray emission itself. \citet{Tagliaferri05} have
also suggested that the {\it patchy shell} model \citep{KPi00b}, where
there are angular inhomogeneities in the outflow, might produce a
sufficiently fast decay if our line of sight is within a {\it hot
spot} in the jet, of angular size $\sim\Gamma^{-1}$, causing a
mini-jet break as the flow is decelerated by the external
medium. However, this would produce $\alpha_1\lesssim p \sim 2-2.5$
(where $p$ is the power law index of the electron energy
distribution), which is significantly lower than the typical observed
values of $3\lesssim\alpha_1\lesssim 5$. Furthermore, this would
require an extreme angular inhomogeneity in the outflow. The {\it
patchy shell} model, however, would naturally produce a series of
bumps and wiggles in the light curve on top of a more moderate
underlying power law flux decay \citep{Fen99, NPG03} (rather than the
observed smooth and very steep decay which later turns into a smooth
and very shallow decay). From all the above, we conclude that while
different mechanisms might still be responsible for the steep early
flux decays in the X-ray afterglows of some GRBs, emission for {\it
naked GRBs} is the most promising mechanism, and is likely at work in
most cases.

\subsection{The First Break in the Light Curve ($t_{\rm break,1}$)}
\label{tb1}

Between $t_{\rm break,1}<t<t_{\rm break,2}$ there is a very shallow
decay of the flux, with $0.5\lesssim\alpha_2\lesssim 1.0 $. We
interpret the first break, at $t_{\rm break,1}$, as the time when the
slowly decaying emission from the forward shock becomes dominant over
the rapidly decaying flux from the prompt emission at large angles
from our line of sight. This break can generally be
chromatic, if the spectrum of the prompt emission at large angles 
(which corresponds to a larger
frequency range in the local frame compared to the observed frequency
during the prompt emission) has a different spectral slope in the
X-rays than the afterglow emission from the forward shock. Under this
interpretation, we do not expect a break in the optical (or UV, or IR)
light curve at exactly the same time as in the X-rays (except for the
rare cases where by coincidence the spectra of these two distinct
physical regions are similar over such a large range in
frequencies). This prediction could serve as a diagnostic test for our
interpretation.

Out of six GRBs for which $t_{\rm break,1}$ was well determined,
two events (050315 and 050319) show clear evidence for a change of the
spectral slope in the X-ray range, $\beta_{\rm X}$, across the break
(with $\Delta\beta_{\rm x}\equiv\beta_{\rm X,2}-\beta_{\rm X,1}$ of
$-0.5$ and $-0.9$, respectively).\footnote{These two GRBs also have a
rather steep early decay with $\alpha_1\approx 4$, which supports the
interpretation of {\it naked GRB} emission.} In the other four cases,
while there is no evidence for a change in $\beta_{\rm X}$ across the
break, such a change can only be constrained to $|\Delta\beta_{\rm
X}|\lesssim 0.3$. Thus we consider the observed behavior of
$\beta_{\rm X}$ across $t_{\rm break,1}$ to be broadly consistent with
our interpretation, in which $|\Delta\beta_{\rm X}|$ is not expected
to always be very large, and can in many cases be rather modest.

The fact that the sharply decaying flux from the prompt emission
initially (at $t<t_{\rm break,1}$) dominates over the emission from
the external shock, suggests that either (i) the prompt emission
dissipates and radiates most of the initial energy in the outflow,
leaving a much smaller energy in the external shock, or (ii) the
energy that is dissipated in the prompt emission (i.e. the kinetic
energy that is converted to internal energy) is comparable to that in
the forward shock, but the fraction of that energy which is radiated
in the observed band is much larger for the prompt emission.  The
latter is relevant for the internal shocks model, in which at most
about half (and typically much less) of the initial kinetic energy is
expected to be converted to internal energy in the internal shocks,
while most of the remaining energy (which is expected to be close to
the original energy) is converted to internal energy in the external
shock. The emission from the forward shock peaks at the deceleration
time (when the ejecta slow down significantly and most of their energy
is transfered to the forward shock), $t_{\rm dec}$, which is
comparable to the duration of the GRB, $T_{\rm GRB}$, for a mildly
relativistic reverse shock, so that a comparable radiative efficiency
would lead to a comparable bolometric luminosity (assuming a similar
fraction of the internal energy goes to electrons and is radiated
away).  Thus, the larger flux from the internal shocks suggests that a
higher fraction of the internal energy is converted into radiation in
the observed band. The high efficiency that is required from the
prompt gamma-ray emission is further discussed in \S \ref{alpha2}.

\subsection{Intermediate Shallow Flux Decay ($\alpha_2$)}
\label{alpha2}

\begin{figure}[ht]
\setlength{\unitlength}{1.0in}
\begin{picture}(6,6)
\put (-0.02,0.0) {\centerline{\includegraphics[width=1.0\textwidth]{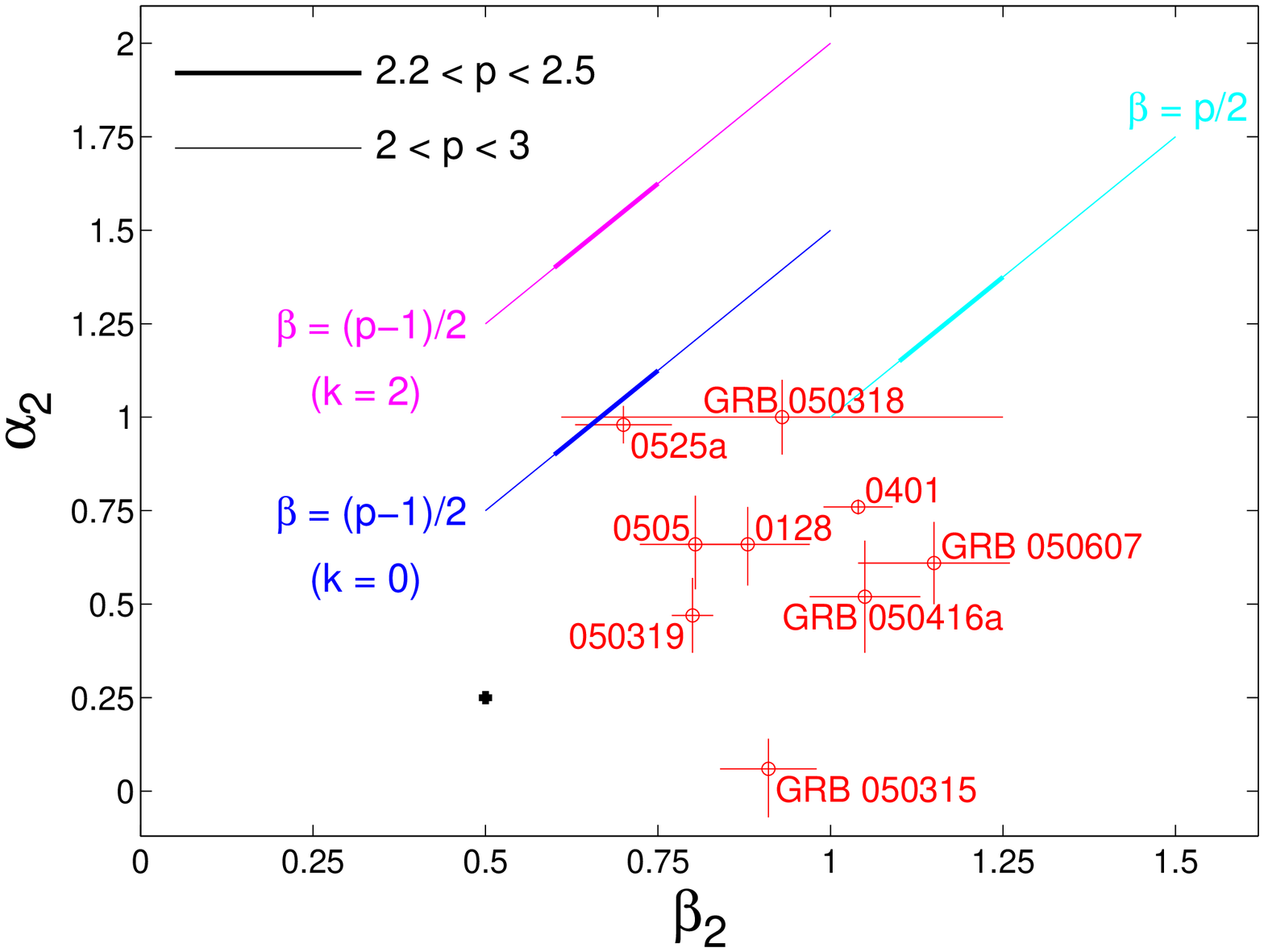}}}
\end{picture}
\caption{\label{alpha2_beta2}The values of $\alpha_2$ and $\beta_2$
  for the GRBs in our sample for which $t_{\rm break,2}$ could be
  determined, as well as the values expected for an adiabatic
  evolution of a spherical afterglow shock \citep{SPN98,GS02}. In most
  relevant power law segments of the spectrum both $\alpha$ and
  $\beta$ depend on $p$, and therefore we drew a thick (thin) line
  corresponding to the range $2.2<p<2.5$ ($2<p<3$) which is both
  typically inferred from GRB afterglows, and is preferred on
  theoretical grounds. The cross at $\alpha_2=1/4$ and $\beta_2=1/2$
  corresponds to the fast cooling power law segment of the spectrum,
  $\nu_c<\nu<\nu_m$, where both $\alpha$ and $beta$ are independent of
  $p$ and $k$.}
\end{figure}

In most cases $\alpha_2$ is too small ($0.5\lesssim\alpha_2\lesssim
1.0$) to be reasonably accounted for with an adiabatic evolution of the
forward shock with a constant energy \citep[][see also \S
  \ref{tb2_alpha3}]{SPN98,GS02}, while radiative losses would only
cause a steeper flux decay.
Figure~\ref{alpha2_beta2} demonstrates this by showing 
the observed values of $(\alpha_2,\beta_2)$ for events for which 
$t_{\rm break,2}$ could be determined, along with the values expected 
from a spherical external shock with a constant energy.  For those
cases where neither adiabatic evolution or radiative losses can
explain the slopes we must instead assume gradual energy
injection during this part of the X-ray light curve\footnote{
Some of the GRB decay curves listed in Table 3 are steep enough to be
consistent with a spherical blast-wave model, but we believe that the more
natural interpretation is to use the same phenomenology for all cases.
We also draw the reader's attention to \citep{Zhang05} and \citep{Panaitescu05}
who make a similar analysis of a sub-sample of nine of the 
27 Swift light curves that are included in this paper, and also provide
alternative explanations for the intermediate shallow flux decay phase 
($\alpha_2$).
Only two of the GRBs reported here (GRB 050318 and 050505), for which the
break at $t_{\rm break,2}$ is monitored, would be consistent with such 
alternative interpretations, and we have chosen the simpler 
characterization.},
which can take
place in two main forms: (i) toward the end of the burst the Lorentz
factor $\Gamma$ of the outflow that is being ejected decreases with
time, forming a smooth distribution of ejected mass as a function of
its Lorentz factor, $M(>\Gamma)$, and its corresponding energy,
$E(>\Gamma)$. In this picture $\Gamma$ increases with radius $R$ and
material with Lorentz factor $\Gamma$ catches up with the forward
shock when the Lorentz factor of the forward shock, $\Gamma_f$, drops
slightly below $\Gamma$ \citep{RM98,SM00,RMR01}, resulting in a smooth
and gradual energy injection into the afterglow shock. (ii) An
alternative scenario for the energy injection is that the central
source remains active for a long time
\citep{RM00,mwh01,RR04,LR04}. Here the ejected outflow has a Lorentz
factor, $\Gamma_i$, that is much larger than that of the forward shock
when it catches-up with it, $\Gamma_i\gg\Gamma_f$.  This leads to a
more highly relativistic reverse shock (with a Lorentz factor
$\Gamma_r\sim\Gamma_i/2\Gamma_f\gg 1$) compared to scenario (i) where
the reverse shock is only mildly relativistic, thus resulting in a
significantly different emission from the reverse shock (which becomes
similar to that from the forward shock for $\Gamma_r\sim\Gamma_f$,
assuming a similar composition and similar micro-physical parameters
in both shocks).

Scenario (ii) requires the central engine to remain active for a very
long time, up to $t_{\rm break,2}$, which is in many cases several
hours (see Table 1 and Figure \ref{tbreaks}). Interestingly enough,
the X-ray flares in the early afterglow light curve of some GRBs also
suggest that the central source remains active for hours after the GRB
(see \S \ref{flares}). The main difference is that scenario (ii)
requires both smooth and continuous (rather than episodic) energy
injection by the source at late times, and that it also requires most
of the energy to be injected at late times, hours after the GRB.

Below, we assume for simplicity that the emission in the X-ray range
is dominated by the forward shock, rather than by the reverse shock,
which is typically expected to be the case.

In scenario (i), the power law flux decay of the X-ray afterglow
suggests power law dependences of $M(>\Gamma)\propto\Gamma^{-s}$ and
$E(>\Gamma)\propto\Gamma^{1-s}$. In order to affect the forward shock
dynamics, slow down its deceleration, and cause a shallower flux
decay, the total energy in the afterglow shock should gradually
increase with time.\footnote{This is valid also when there are
radiative losses, in which case an increase with time in the energy of
the afterglow shock would require a faster energy injection rate,
corresponding to a higher minimal value of $s$, compared to the
requirement $s>1$ for the adiabatic case. For simplicity we neglect
radiative losses in the following analysis.} This implies that the
total injected energy, $E_i(t)$, must first exceed the initial energy,
$E_0$, in the afterglow shock before it significantly affects its
dynamics. Therefore, the flattening of the light curve would start at
$t_i$ for which $E_i(t_i)\sim E_0$. Neglecting radiative losses we
have $E(t)=E_0+E_i(t)$, so that $E(t)\approx E_i(t)$ for
$t>t_i$. Furthermore, $E_i(t)$ should gradually increase with time,
implying $s>1$. For simplicity spherical symmetry is assumed below,
but the results are also valid for a uniform jet (viewed from within
its aperture) as long as the Lorentz factor exceeds the inverse of the
jet half-opening angle, and when the energy is replaced by the
isotropic equivalent energy. For any given power law segment of the
spectrum we have $F_\nu\propto E^b t^{-\alpha_{\rm ad}}$ where
$\alpha_{\rm ad}$ is the temporal decay index for an adiabatic shock
evolution (with no energy injection or radiative losses) which is
given in Eq. \ref{alpha3}, while $E(t>t_i)\propto t^a$ with
$a=(s-1)(3-k)/(7+s-2k)$ for energy injection with an external density
profile $\rho_{\rm ext}=Ar^{-k}$ \citep{SM00}. For the relevant power
law segments of the spectrum,
\begin{equation}\label{b}
b=\left\{\matrix{3/4&=&3\beta/2 & \nu_c<\nu<\nu_m & (k<3)\ ,\cr\cr
(3+p)/4 &=& (\beta+2)/2 & \nu_m<\nu<\nu_c & (k=0)\ , \cr\cr (1+p)/4 &=&
(\beta+1)/2 & \nu_m<\nu<\nu_c & (k=2)\ , \cr\cr (2+p)/4 &=& (\beta+1)/2 &
\ \nu>\max(\nu_m,\nu_c) & (k<3)\ ,}\right.
\end{equation}
\citep{GS02}. The increase in the temporal decay index across the
break at $t_{\rm break,2}$ is $\Delta\alpha\equiv\alpha_3-\alpha_2 =
a\,b$. Thus, we can obtain the power law index of the energy
injection, $s$, as a function of $\beta$ [or $b(\beta)$] and
$\Delta\alpha$ which may be directly measured from observations,
\begin{equation}\label{s}
s=1+\frac{2(4-k)\Delta\alpha}{(3-k)b(\beta)-\Delta\alpha}\ .
\end{equation}

One can determine the power law segment in which the X-ray band is
located, and thus the appropriate expression for $b(\beta)$ (see
Eq. \ref{b}), from the relations between $\alpha_3$ and $\beta_3$ (see
\S \ref{tb2_alpha3}). Figure \ref{alpha3_beta3} shows the values of
$(\alpha_3,\beta_3)$ for the events for which $t_{\rm break,2}$ could
be determined, along with the expected relations for the potentially
relevant power law segments of the spectrum. There are nine such
events, and they all fall reasonably close to these relations.

Table 3 gives the derived values of $s$,
for the nine GRBs in which the
Swift observations show a shallow decay segment, and hence have a
$\Delta\alpha$ indicating energy injection.  There are eight other GRBs
which were not observed by Swift to have a shallow segment, 
and so do not have a $\Delta\alpha$,
but we cannot rule out that energy injection might have occurred during
times when Swift could not observe these GRBs.  
In this scenario it should
be noticed that $s>1$ so that the value of $a$ is bound within the
range $0<a<(3-k)$, and correspondingly
\begin{equation}\label{alpha_max}
0<\Delta\alpha<\Delta\alpha_{\rm max}=(3-k)b(\beta)\ ,
\end{equation}
where $\Delta\alpha$ approaches $\Delta\alpha_{\rm max}$ for $s\gg
1$. The limits on the possible values of $\Delta\alpha$ in scenario
(i) are more constraining for a stellar wind environment ($k=2$) for
which $0<\Delta\alpha<b(\beta)\sim 1$, compared to a uniform external
density ($k=0$).\footnote{Note, however, that the values for $s$ have 
been derived
under the assumption, $k=0$, (an homogeneous medium).  Because the cooling
frequency is below the X-ray, we cannot distinguish between the $k=0$
and $k=2$ case (wind-like medium).  If $k=2$, then a different value 
for $s$ would be determined.}
We find only the high redshift GRB 050505 not to be
compatible with this constraint.\footnote{The reader should, however,
keep in mind that for this event the determination of $t_{\rm
break,2}$ is uncertain given the intrinsic curvature of the afterglow
lightcurve.}

On the other hand, in order to reproduce the observed power law decay
of the X-ray flux in scenario (ii), the (kinetic) luminosity of the
central source should be a power law in the observer frame time $t_{\rm
lab}$ (for which $R \approx ct_{\rm lab}$), $L\propto t_{\rm
lab}^q$. In this case $\Gamma\propto R^{-(2-q-k)/2(2+q)}\propto
t^{-(2-q-k)/2(4-k)}$ \citep{BM76} and $E\propto t^{q+1}$,
i.e. $a=q+1$. Therefore, for scenario (ii) we have the relatively
simple relation,
\begin{equation}\label{q}
q=\frac{\Delta\alpha}{b(\beta)}-1\ . 
\end{equation}

Table 3 gives the required values of $q$ for various bursts in our
sample. It is interesting to note that in scenario (ii) there is no
upper bound on the value of $a$ or on the values of
$\Delta\alpha=a\,b(\beta)$, but only a trivial lower limit ($a>0$ and
$\Delta\alpha>0$). This is in contrast with scenario (i) where
$\Delta\alpha$ has an upper limit of $\Delta\alpha_{\rm
max}=(3-k)b(\beta)$ (see Eq. \ref{alpha_max}). Therefore, if
$\Delta\alpha$ exceeds $\Delta\alpha_{\rm max}$ for some GRB, this
could be explained only by scenario (ii), and not by scenario (i).
This can potentially serve as a diagnostic method for distinguishing
between these two types of energy injection into the forward shock.

In both scenarios discussed above, the total amount of injected energy
must increase with time (and exceed the initial energy in the
afterglow shock) to effect the dynamics of the afterglow shock and
cause a shallower flux decay. In scenario (ii), this implies $q>-1$,
which is not a trivial requirement, and is hard to produce in many GRB
progenitor models. For example, in the collapsar model the late time
accretion rate due to fallback is expected to scale with time as
$\dot{M}_{\rm acc}\propto t_{\rm lab}^{-5/3}$ \citep{mwh01}, which for
a roughly constant efficiency, $\eta$, implies $L=\eta\dot{M}_{\rm
acc}c^2\propto t_{\rm lab}^{-5/3}$ and $q=-5/3$. For the magnetar
model \citep{zhang,rosswog,usov,dai}, $L$ is initially constant while
after the newly born neutron star spins down significantly, $L\propto
t_{\rm lab}^{-2}$, i.e. $q=-2$.  Thus, neither of these models can 
naturally explain the flatter flux decays at $t_{\rm break,1}<t<t_{\rm
break,2}$ due to late time energy injection from the source (see Table
3).

Regardless of the exact details of the energy injection, we can
constrain the factor $f$ by which the energy of the afterglow shock
was increased due to the energy injection [$f=(E_0+E_{\rm
injected}-E_{\rm radiated})/E_0$ where $E_{\rm injected}=E_i(t_f)$].
If the energy injection lasted between $t_i$ and $t_f$, it would cause
a flux increase by a factor of $(t_f/t_i)^{\Delta\alpha}$ compared to
the hypothetical case of no energy injection (or radiative losses),
corresponding to $f=(t_f/t_i)^{\Delta\alpha/b(\beta)}$. While $t_f$ is
identified with $t_{\rm break,2}$, we do not know the exact value of
$t_i$. We do, however, know that $t_i<t_{\rm break,1}$, and
$t_i\gtrsim T_{\rm GRB}$, which provide the following constraints on
$f$,
\begin{equation}\label{f}
\left(\frac{t_{\rm break,2}}{t_{\rm break,1}}\right)^{\Delta\alpha/b(\beta)}
    <f\lesssim\left(\frac{t_{\rm break,2}}{T_{\rm
    GRB}}\right)^{\Delta\alpha/b(\beta)}\ .
\end{equation}

The energy in the afterglow at late times (later than several hours
and therefore at $t>t_{\rm break,2}$) is typically estimated to be
comparable to or smaller than that in the prompt gamma-ray emission
\citep{PK02,L-RZ04}, even when correcting for radiative losses from
the afterglow shock at early times, implying a high efficiency for the
prompt emission, which is $\gtrsim 50\%$ in most cases. This is in
particular a serious problem for the internal shocks model, where it
is hard to reach such high efficiencies in converting the bulk kinetic
energy of the outflow to the observed gamma-rays
\citep{Kumar99,GSW01}. The energy injection interpretation implies
that most of the energy in the afterglow shock at late times was
either (i) originally in material with an initial Lorentz factor
$\Gamma < 10^2$ which could therefore have not contributed to the
prompt gamma-ray emission 
\citep[due to the compactness problem;][]{LS01,Piran99} or (ii) injected at late times,
after the prompt gamma-ray emission was over.

This requires the prompt gamma-ray emission to be significantly more
efficient than previous estimates, where $E_\gamma/E_0$ increases by a
factor of $f\gtrsim 4$ (see Table 3). Furthermore, we find that the
energy of the afterglow shock increases by a factor of $f$ when also
taking into account radiative losses, while most previous estimates of
$E_\gamma/E_0$ included the radiative losses assuming that they
decrease the energy of the afterglow shock by a factor of $\sim 3$ or
more \citep[][Panaitescu, private communication]{PK01a,PK01b,L-RZ04}.
Therefore, the correction for $E_\gamma/E_0$ compared to such
previous estimates would be even larger (by a factor of $\sim
3f$). The efficiency of the prompt gamma-ray emission is usually
defined as $\epsilon_\gamma\equiv
E_\gamma/(E_\gamma+E_0)=1/(1+E_0/E_\gamma)$. Thus, previous estimates
which typically gave $E_\gamma/E_0\gtrsim 1$ and
$\epsilon_\gamma\gtrsim 50\%$ would now, with a correction factor of
$\sim 3f\gtrsim 10$, imply $\epsilon_\gamma\gtrsim 90\%$. This poses
severe requirements for theoretical models.

While we have concentrated on energy injection into the forward shock as
the explanation for the early flat decay phase in the X-ray afterglows,
there are also alternative explanations. Here we briefly mention a few
such alternatives. A very shallow power law index of the
electron energy distribution, $p<2$, might explain the shallow decay phase
in the few cases where the spectral and temporal indexes are consistent
with this picture (i.e. $\beta_2 = p/2 < 1$ and $\alpha_3\approx p$ if the
break at $t_{\rm break,2}$ is attributed to a collimated outflow where the
edge of the jet becoming visible at this time). Another possible
explanation that might be at work in at least some cases is a viewing
angle slightly outside the region in the GRB jet with a prominent
afterglow emission \citep{EicGra05}. A more radical explanation
might be found in the context of the cannonball model \citep{Dado05}.

\subsection{Second Break in the Light Curve ($t_{\rm break,2}$) into a
  Steeper Flux Decay ($\alpha_3$)}
\label{tb2_alpha3}

When energy injection ends, at $t>t_{\rm break,2}$, an adiabatic
evolution of the forward shock at a constant energy follows, producing
a somewhat steeper decay slope, $\alpha_3$ \citep{SPN98,GS02}. The
relations between the temporal and spectral indices for the power law
segments of the spectrum that might be relevant in the X-rays are
\begin{equation}\label{alpha3}
\alpha=\left\{\matrix{1/4&=&\beta/2 & \nu_c<\nu<\nu_m & (k<3)\ ,\cr\cr
3(p-1)/4 &=& 3\beta/2 & \nu_m<\nu<\nu_c & (k=0)\ , \cr\cr 
(3p-1)/4 &=& (3\beta+1)/2 & \nu_m<\nu<\nu_c & (k=2)\ , \cr\cr 
(3p-2)/4 &=& (3\beta-1)/2 &
\ \nu>\max(\nu_m,\nu_c) & (k<3)\ .}\right.
\end{equation}

In this picture $t_{\rm break,2}$ corresponds, for the two scenarios
described in \S \ref{alpha2}, respectively, to (i) the time when the
energy injection to the forward shock ends, i.e. when the Lorenz
factor of the forward shock drops to slightly below the minimal
Lorentz factor, $\Gamma_{\rm min}$, of the ejecta which carry
significant energy, or (ii) the time when the central source becomes
inactive. Under both of the energy injection scenarios, the second
break (at $t_{\rm break,2}$) should be achromatic, as long as the
emission before the break was dominated by the forward shock rather
than by the reverse shock. If the emission before the break is
dominated by the reverse shock, then there should be a brief period of
fast decay of the flux (from the reverse shock), as the supply of
newly shocked outflow ends, and the existing shocked outflow cools
adiabatically (and radiatively). This short phase ends once the
emission becomes dominated by the forward shock. During this short
intermediate period the light curve could show chromatic behavior.

Assigning a single Lorentz factor ($\Gamma$) to a given observed time
is not posssible, as photons from a wide range of radii and Lorentz factors
reach the observer simultaneously. One can parameterize
$t/(1+z)=R/C\Gamma^2c$, where the uncertainty is put into the value of
the parameter $C$. Using the \citet{BM76} self-similar solution as a
guide, and evaluating the Lorentz factor just behind the shock, this
gives
\begin{equation}\label{gamma_BM}
\Gamma = \left[\frac{(17-4k)E(1+z)^{3-k}}{16C^{3-k}\pi
c^{5-k}t^{3-k}}\right]^{1/2(4-k)}\ .
\end{equation}
Estimating the typical Lorentz factor for a given observed
time as that just behind the shock at the radius corresponding to the
outer edge of the afterglow image \citep{GS02} gives $C=4(4-k)/(5-4)$ and in
turn,
\begin{equation}\label{gamma1}
\Gamma(t) = \left\{\matrix{6.68(E_{52}/n_0)^{1/8}[t_{\rm
days}/(1+z)]^{-3/8} & \quad (k=0)\ , \cr\cr
4.90(E_{52}/A_*)^{1/4}[t_{\rm days}/
(1+z)]^{-1/4} & \quad (k=2)\ ,}\right.
\end{equation}
while simply parameterizing $C=4C_4$ gives
\begin{equation}\label{gamma2}
\Gamma(t) = \left\{\matrix{6.14\,C_4^{-3/8}(E_{52}/n_0)^{1/8}[t_{\rm
days}/(1+z)]^{-3/8} & \quad (k=0)\ , \cr\cr
4.43\,C_4^{-1/4}(E_{52}/A_*)^{1/4}[t_{\rm days}/
(1+z)]^{-1/4} & \quad (k=2)\ ,}\right.
\end{equation}
where $t_{\rm days}=t/(1\;{\rm day})$, $n=n_0\;{\rm cm^{-3}}$ is the
external number density for $k=0$, $A_*=A/(5\times 10^{11}\;{\rm
g\;cm^{-1}})$ for $k=2$, and $E_{52}=E/(10^{52}\;{\rm erg})$.  In
scenario (i) one may estimate $\Gamma_{\rm min}\approx 2\Gamma(t_{\rm
break,2})$, using equations \ref{gamma1} or \ref{gamma2} for
$\Gamma(t)$. Typical values are $15\lesssim\Gamma_{\rm min}\lesssim
50$ for $k=0$ and $10\lesssim\Gamma_{\rm min}\lesssim 20$ for $k=2$
[for $E_{52}/n_0\sim 1$, $E_{52}/A_*\sim 1$ and $0.2\lesssim t_{\rm
break,2}/(10^4\;{\rm s})\lesssim 4$].

Optical afterglows typically show a jet break at $t_{\rm jet}$ which
can range from several hours to weeks, and typically occurs after
$t_{\rm break,2}$. Thus, it might be possible in some cases to see the
jet break at $t_{\rm jet}>t_{\rm break,2}$, as might be the case for
GRB 050315 \cite{Vaughan05}.

\subsection{X-ray Flares in the Early Afterglow}
\label{flares}

The early X-ray light curves obtained with {\it Swift} XRT often show
flares (see {\it lower panel} of Fig. \ref{F_X(t)}). The most
prominent flare so far was in GRB 050502B, where the flux increased by
a factor of $\sim 500$. Some of these flares have very sharp temporal
features where the flux changes significantly on time scales $\Delta
t\ll t$ \citep{Burrows05}. Most flares have a very steep rise and
decay (with very large temporal rise/decay indices when fitted to a
power law). When the flare is bright enough to follow its spectral
evolution, its hardness ratio evolves during the flare, and its
spectral index is somewhat different from the one associated with the
underlying power law decay of the X-ray light curve before and after
the flare \citep{Burrows05}. Furthermore, the fluxes before and after
the flare lie approximately on the same power law decay, suggesting
that the flare originates from a different physical component than that
responsible for the underlying power law decay.

It is very difficult (nearly impossible under realistic conditions) to
produce very sharp temporal variations of the flux ($\Delta t\ll t$)
with large amplitudes ($\Delta F\gtrsim F$) in the external shock, be
it from refreshed shocks\footnote{Refreshed shocks that occur after
the jet break in the light curve, could produce $\Delta F\gtrsim F$ on
time scales $\Delta t/t$ as small as $\sim 0.15-0.2$ with
\citep{GNP03}, corresponding to the ratio of the radial and angular
times. The X-ray flares, however, typically occur at early times,
before the jet break time, so that we expect $\Delta t/t\sim 1$ for
refreshed shocks.} \citep{KPi00a,GNP03,RMR01}, bumps in the external
medium \citep{Lazzati02,NPG03,Rwinds01} or angular inhomogeneities in
the outflow \citep{Fen99,NPG03}. Therefore, the most likely
explanation for these flares is late internal shocks. This implies
that the central source is still active at relatively late times.

\section{Conclusions}
\label{conc}

\begin{figure}[h]
\setlength{\unitlength}{1.0in}
\begin{picture}(6,7)
\put (0.01,5.0)
     {\centerline{\includegraphics[width=118mm,angle=-90,bb=78 21 572
	   699,clip]{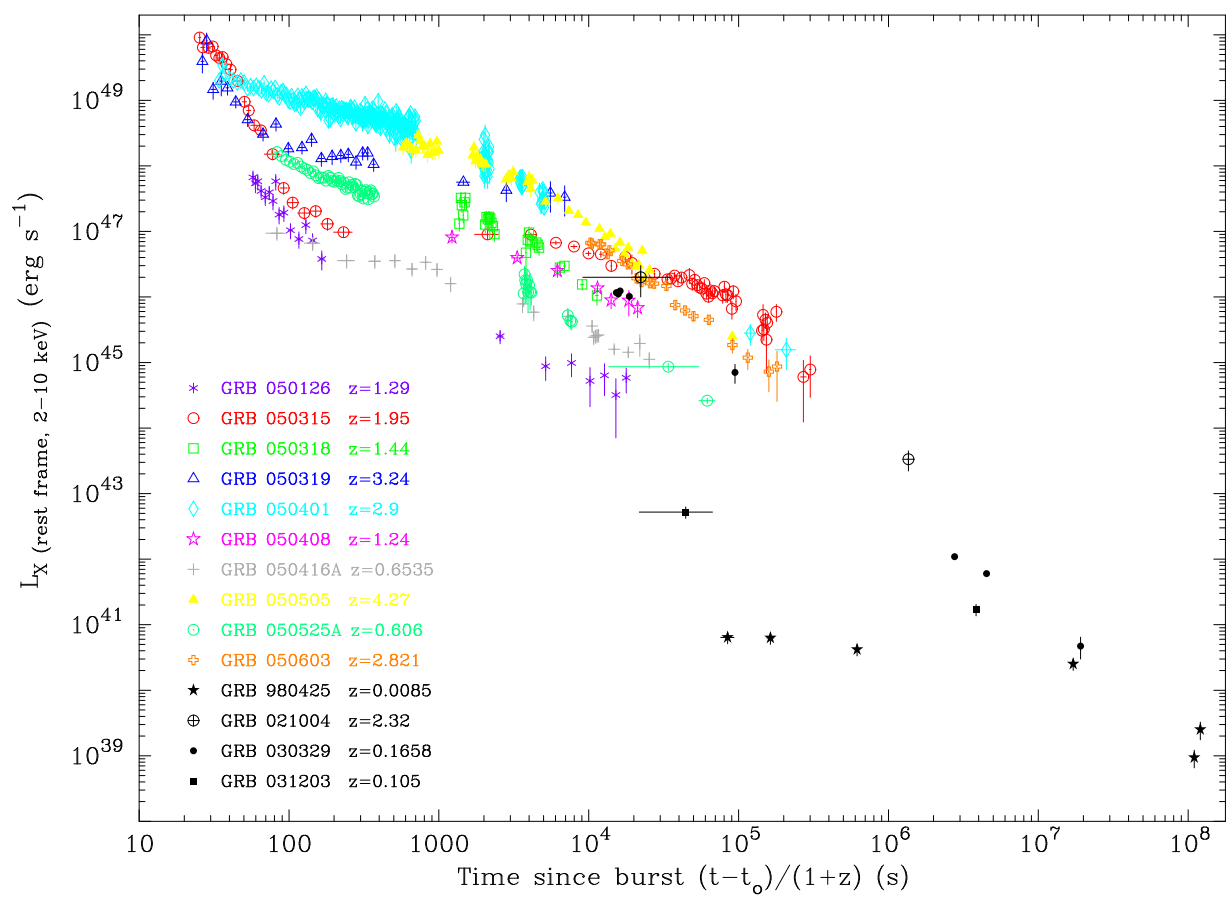}}}
\end{picture}
\caption{\label{L_X(t)}The X-ray luminosity in the range $2-10\;$keV
  as a function of time (both measured in the cosmological rest frame
  of the GRB) for \swift\ GRBs with established redshifts (coloured symbols), plotted
  together with selected earlier events (all in black symbols) from Figure 3 in
  \cite{kouv04}.}
\end{figure}

\begin{figure}
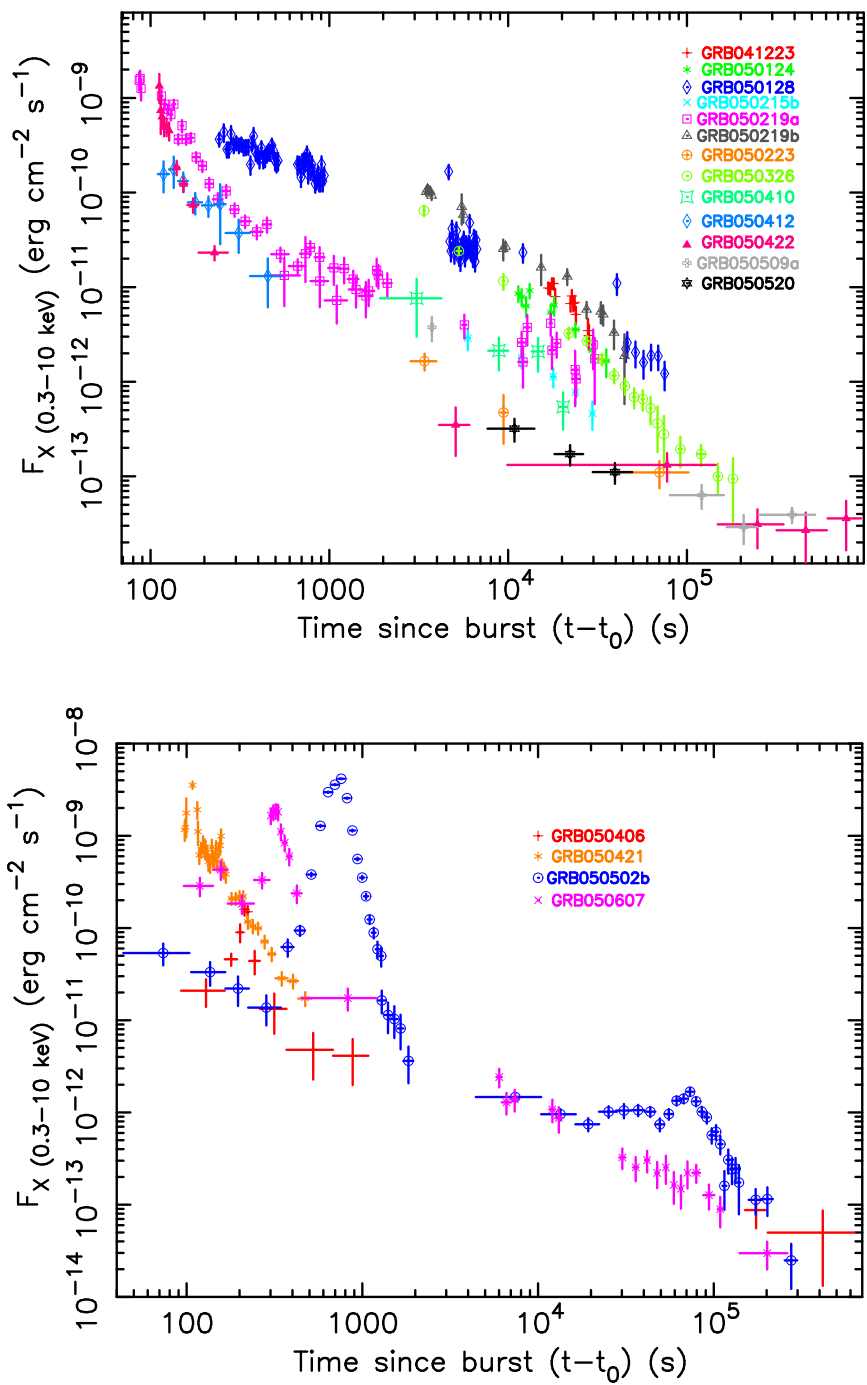

\setlength{\unitlength}{1.0in}
\begin{picture}(6,7.5)
\put (0.0,7.1)
     {\centerline{\includegraphics[width=86mm,angle=-90,bb=78 21 572
	   699,clip]{fluxlc_new_noflare.ps}}}
\put (0.077,3.45) {\centerline{\includegraphics[width=90.4mm,angle=-90,clip]{fluxlc_new_flares.ps}}}
\end{picture}
\caption{\label{F_X(t)}The X-ray flux ($0.3-10\;$keV in the observer
frame) as a function of the observed time, for all \swift\ GRBs
without known redshifts, with ({\it lower panel}) and without ({\it
upper panel}) X-ray flares.}
\end{figure}

\begin{figure}[h]
\setlength{\unitlength}{1.0in}
\begin{picture}(6,7)
     {\centerline{\includegraphics[width=0.9\textwidth]{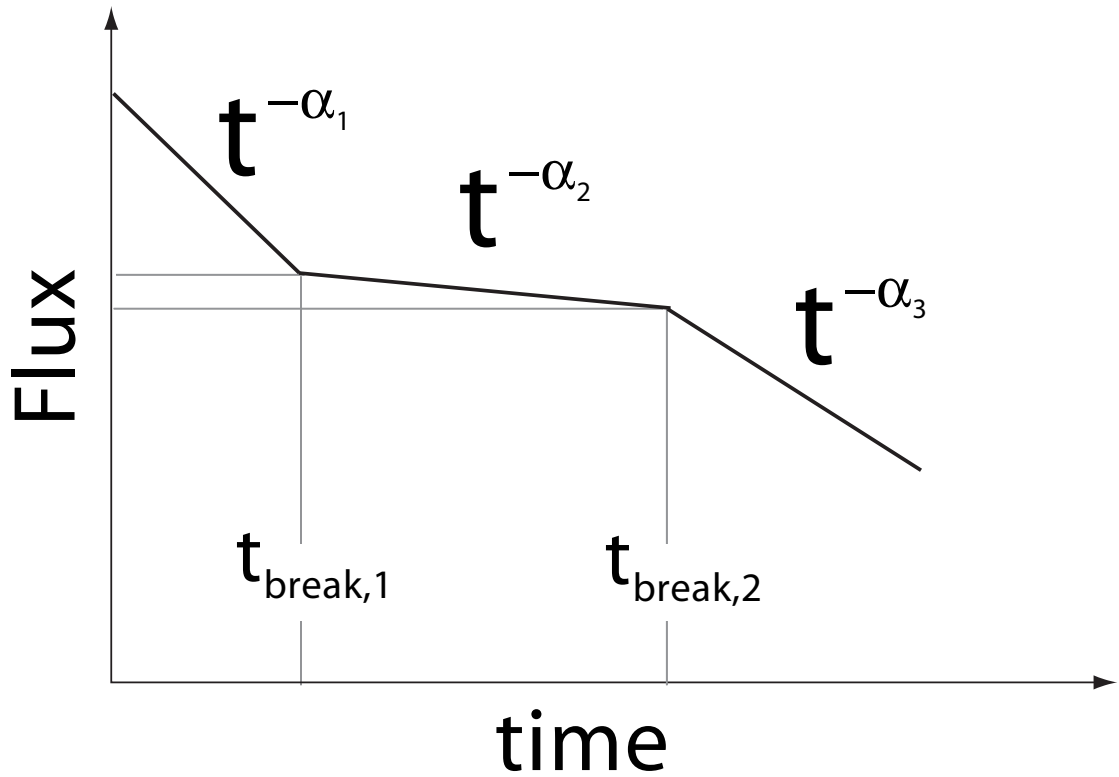}}}
\end{picture}
\caption{\label{diagram} A schematic diagram of the
  canonical behavior of the early X-ray light curve for GRBs observed
  with \swift\ XRT. It consists of three power law segments where
  $F_\nu\propto\nu^{-\beta}t^{-\alpha}$: (i) a fast initial decay with
  $3\lesssim\alpha_1\lesssim 5$, (ii) a very shallow decay with
  $0.5\lesssim\alpha_2\lesssim 1.0$, (iii) a somewhat steeper decay
  with $1\lesssim\alpha_3\lesssim 1.5$. The transition between these
  power law segments occurs at two break times, $t_{\rm break,1}$
  and $t_{\rm break,2}$.}
\end{figure}

We have presented X-ray light curves for 27 GRBs monitored by \swift\
XRT during December 2004 - June 2005. These light curves start as
early as $\lesssim 10^2\;$s after the GRB trigger, and cover up to
four decades in time. The most striking result we obtain is that the
early X-ray light curves show a canonical behavior (see
Figs. \ref{L_X(t)}, \ref{F_X(t)} and \ref{diagram}) which consists of
three power law segments: an initial very steep decay ($F_\nu\propto
\nu^{-\beta}t^{-\alpha}$ with $3\lesssim\alpha_1\lesssim 5$), followed
by a very shallow decay ($0.5\lesssim\alpha_2\lesssim 1.0$), and
finally a somewhat steeper decay ($1\lesssim\alpha_3\lesssim
1.5$). These three power law segments of the early X-ray light curve
meet at two break times, $t_{\rm break,1}\lesssim
500\;$s, and $10^3\;{\rm s}\lesssim t_{\rm break,2}\lesssim
10^4\;$s. All the light curves in our sample are consistent with this
basic picture of a canonical light curve, although in many cases we do
not see all three power law segments, due to limited temporal
coverage.  

\begin{figure}[ht]
\setlength{\unitlength}{1.0in}
\begin{picture}(6,7)
\put (-0.02,0.0) {\centerline{\includegraphics[width=0.95\textwidth]{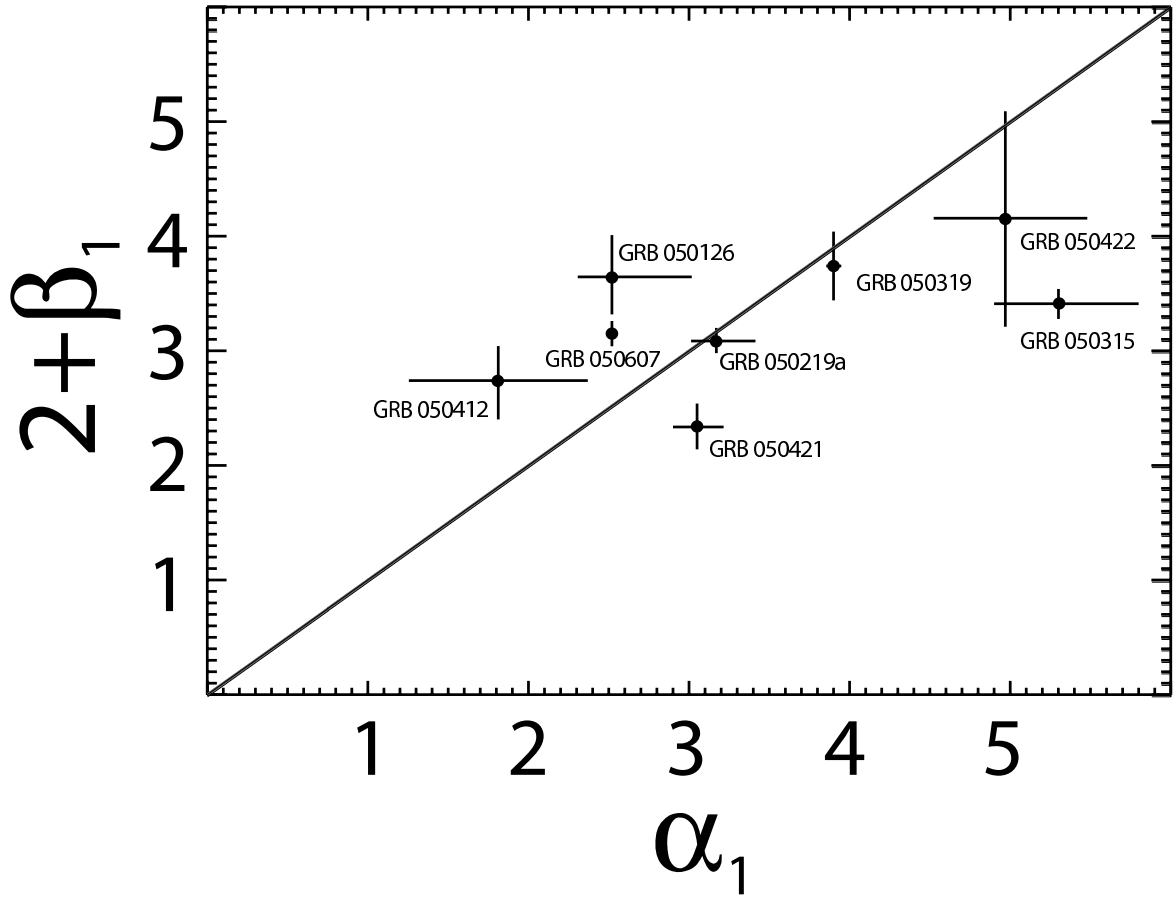}}}
\end{picture}
\caption{\label{alpha1beta1} $\alpha_1$ as a function of
$2+\beta_1$. The solid line gives the theoretical prediction for the
prompt gamma-ray tails emitted at large angles ($\theta>\Gamma^{-1}$)
relative to our line of sight \citep{KP00}.}
\end{figure}

The large variety of behaviors exhibited by afterglows at different
times in their evolution, while clearly
compatible with relativistic fireball models, poses new challenges of
interpretation. We find that the most promising explanation for the
initial fast flux decay ($\alpha_1$) is that it is the tail of the
prompt gamma-ray emission which is emitted from large angles
($\theta>\Gamma^{-1}$) relative to our line of sight
\citep{KP00}. This model produces a sharp flux decay with
$\alpha_1=2+\beta_1$, in rough agreement with observations
(Fig. \ref{alpha1beta1}), while $\alpha_1>2+\beta_1$ might also be
expected for $t/T_{\rm GRB}\lesssim{\rm a\ few}$ (see \S
\ref{alpha1}).

The shallow intermediate flux decay ($\alpha_2$) is most likely caused
by continuous energy injection into the forward shock. This energy
injection is probably due to a decrease in the Lorentz factor,
$\Gamma$, of the outflow toward the end of the prompt GRB, resulting
in a monotonic increase of $\Gamma$ with radius. This outflow
gradually catches up with the afterglow shock, resulting in a smooth
energy injection \citep{SM00}. This picture requires
$E(>\Gamma)\propto\Gamma^{1-s}$ with $s>1$. We have deduced the values
of $s$ from the observed X-ray light curves (see Table 3) and
typically obtain $s\sim 2.5$.

Energy injection could also be caused by a long lasting activity of
the central source, which keeps ejecting significant amounts of energy
in a highly relativistic outflow up to several hours after the
GRB. However, this requires the source luminosity to decay very slowly
with time, $L\propto t_{\rm lab}^q$ with $q>-1$, where most of the
energy is extracted near $t_{\rm break,2}$, i.e. up to several hours
after the GRB.  One might be able to distinguish between these two 
scenarios for energy
injection by the help of early broad band observations, since the
emission from the reverse shock is expected to be different for these
two cases. Furthermore, the change in the temporal index,
$\Delta\alpha$, across the second break in the light curve at $t_{\rm
break,2}$ is bounded in the first scenario (see Eq. \ref{alpha_max})
but not in the second scenario. In all the GRBs in our sample for
which it could be tested (perhaps with one exception, GRB 050505), the
value of $\Delta\alpha$ falls within the allowed range for the first
scenario.

\begin{figure}[ht]
\setlength{\unitlength}{1.0in}
\begin{picture}(6,6)
\put (-0.02,0.0) {\centerline{\includegraphics[width=1.0\textwidth]{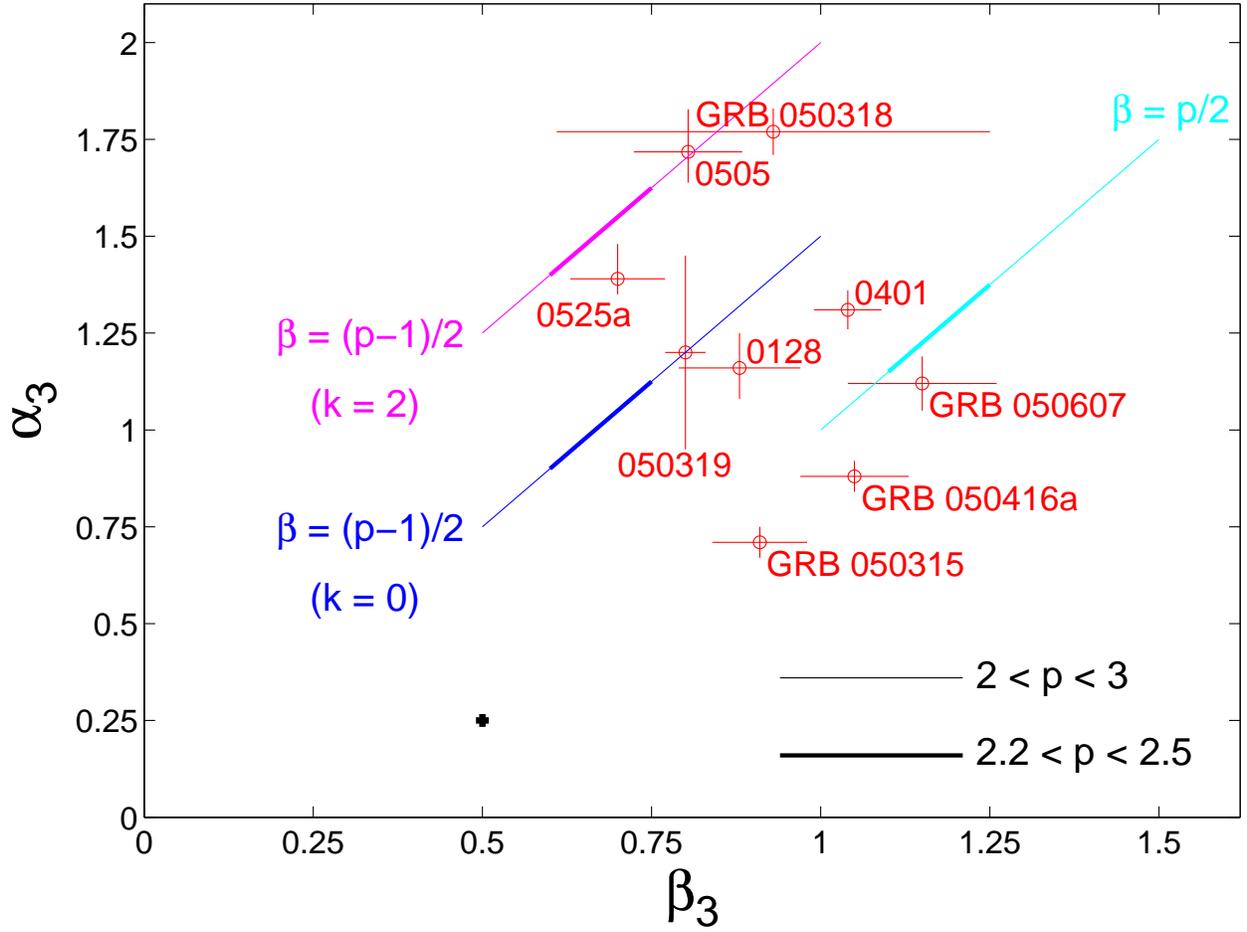}}}
\end{picture}
\caption{\label{alpha3_beta3}The values of $\alpha_3$ and $\beta_3$
  for the GRBs in our sample for which $t_{\rm break,2}$ could be
  determined.  Other symbols as in Fig \ref{alpha2_beta2}.}
  
\end{figure}

The third power law segment of the light curve ($\alpha_3$) is most
likely the well known afterglow emission from a spherical adiabatic
external shock \citep{SPN98,GS02}. The observed values of the temporal
index ($\alpha_3$) and the spectral index ($\beta_3$) are consistent
with this interpretation (see Fig.~\ref{alpha3_beta3}).

In some cases flares are seen on top of the basic canonical light
curve that is illustrated in Fig.~\ref{diagram}, as can be seen in the
{\it lower panel} of Fig.~\ref{F_X(t)}. These flares are most likely
caused by internal shocks within the outflow that is ejected from the
central source at late times (very close to the time when these flares
are seen). This implies that the central source quite often remains
active for hours after the GRB.

We find evidence for a change in the spectral slope across the first
break in the light curve ($t_{\rm break,1}$) in two out of six cases
for which we could determine $t_{\rm break,1}$. This is consistent
with our interpretation in which the first break occurs when the
slowly decaying emission from the forward shock becomes dominant over
the steeply decaying tail emission of the prompt GRB from large angles
with respect to our line of sight. Since these two components arise
from physically distinct regions, their spectrum would generally be
different. We also find no evidence for a change in the spectral slope
across the second break in the light curve ($t_{\rm break,2}$). This
also agrees with our interpretation that this break is caused by
the end of the energy injection into the forward shock, as long as the
emission before the break (at $t_{\rm break,1}<t<t_{\rm break,2}$) is
dominated by the forward shock (rather than by the reverse shock),
which is typically expected to be the case in the X-ray band.

Finally, the interpretation of the shallow intermediate flux decay as
caused by energy injection, implies that the energy in the afterglow
shock at late times (more than several hours) is larger than that at
the deceleration time by a factor of $f\gtrsim 4$ (see Eq. \ref{f} and
Table 3). As discussed at the end of \S \ref{alpha2}, this requires
the prompt gamma-ray emission to be extremely efficient, and typically
convert $\gtrsim 90\%$ of the total energy in the highly relativistic
outflow (with $\Gamma\gtrsim 10^2$) that is ejected during the GRB
itself into the observed gamma-rays. If a significant fraction of the
radiated energy goes to photon energies above the observed range, the
efficiency requirements of the prompt emission become even more
severe.

\acknowledgments The authors acknowledge support from ASI, NASA and
PPARC, and benefits from collaboration within the EC FP5 Research
Training Network ``Gamma-Ray Bursts -- An Enigma and a Tool''. C. K.,
S. K. P. and J. G. acknowledge the hospitality of the Institute of
Advanced Study in Princeton, during the preparation of this
paper. This research was supported by US Department of Energy under
contract number DE-AC03-76SF00515 (J.G.), by NASA through a Chandra
Postdoctoral Fellowship award PF3-40028 (E. R.-R.), and by NASA contract
NAS5-00136 (J. N., D. G., D. B., A. F., J. K.).

{}

\newpage

\newpage

\newpage

\newpage

\newpage

\newpage

\newpage

\newpage

\newpage

\clearpage

\begin{deluxetable}{lcccccccc}
\tabcolsep=1pt
\scriptsize
\tablewidth{0pt}
\tablecaption{TEMPORAL PARAMETERS OF \swift/XRT GRBs}
\tablehead{
\colhead{Burst} &  \colhead{$\alpha_1$} &  \colhead{t$_{\rm break,1}$} &
\colhead{$\alpha_2$} & \colhead{t$_{\rm break,2}$} & \colhead{$\alpha_3$}
&
\colhead{$\beta_{\rm 1,x}$} & \colhead{$T_{\rm burst}^b$} &
\colhead{XRT onset} \\
\colhead{} &  \colhead{} &  \colhead{s} &  \colhead{} &  \colhead{s} &
\colhead{} &  \colhead{$\beta_{\rm 2,x}$} & \colhead{} & \colhead{s} 
}
\startdata
041223 &                 &                &
      &     &    1.72$\pm$0.20    &
1.04$\pm$0.17 & 14:06:18  & $16661$  \\
050124 &                &                 &    &
&    1.58$\pm$0.11              & 0.66$\pm$0.20 & 11:29:48 &
$11113$ \\
050126 &    2.52$^{+0.50}_{-0.22}$    & 424$^{+561}_{-120}$ &
1.00$^{+0.17}_{-0.26}$    &            &            &
1.64$\pm$0.37 & 12:00:55  & $131$ \\
050128 &      &    & 0.66$^{+0.10}_{-0.11}$    & 1724$^{+937}_{-565}$ &
1.16$^{+0.09}_{-0.08}$    &     0.88$\pm$0.09 & 04:19:56  & $107.7$ \\
050215B &                &              &
0.75$\pm$0.27      &            &            & 0.44$\pm$0.42 &
02:33:44 & $>6000$ \\
050219A &    3.17$^{+0.24}_{-0.16}$ & 332$^{+26}_{-22}$    &
0.75$^{+0.09}_{-0.07}$ &             &            &
1.09$\pm$0.11 & 12:39:56 &  $92.4$ \\
050219B &            &                &   &
&    1.20$\pm$0.09            & 1.14$\pm$0.09 & 21:04:49 &
$3129$ \\
050223 &                &              &
0.99$^{+0.15}_{-0.12}$    &            &            &
0.75$\pm$0.19 & 03:09:10 & $2874$ \\
050315 & 5.3$^{+0.5}_{-0.4}$ & 400$\pm$20  & 0.06$^{+0.08}_{-0.13}$      &
12000$\pm$400 &    0.71$\pm$0.04        &
$1.42\pm 0.12$ & 20:59:40 & $83$ \\
   &   &   &   &   &   &  $0.91\pm 0.07^c$ &   &   \\
050318 &     &   & 1.00$\pm$0.10 & 10000$\pm$100    & 1.77$\pm$0.06    &
0.93$\pm$0.32 & 15:44:37 & $3276.8$  \\
050319 &    3.9$\pm$0.05        & 370$\pm$15    & 0.47$\pm$0.10    &
40000$\pm$300 & 1.2$\pm$0.25 & 1.74$\pm$0.3 &
09:29:04 & $87.09$ \\
   &   &   &   &   &   &  0.8$\pm$0.03 &   &   \\
050326 &                &                &     &
& 1.60$\pm$0.06 & 0.80$\pm$0.27 & 09:53:46 & $3259$ \\
050401 &     &  &  0.76$\pm$0.02    & 5518$^{+1149}_{-1043}$ &
1.31$\pm$0.05     & 1.04$\pm$0.05 & 14:20:08 & $127$ \\
050406$^e$ &                &            &  & &  & $1.32\pm0.15$ & 15:58:48 & $86.6$ \\
050408 &                &            &  &
&    0.83$\pm$0.04            & 1.16$\pm$0.11 & 16:22:50 &
$2547.1$ \\
050410 &                &            &  &
&    1.15$\pm$0.10            & 1.05$\pm$0.28 & 12:14:34 &
$1921.6$ \\
050412 & $1.81^{0.57}_{0.47}$  & &      &            &            &
0.74$\pm$0.32 & 05:43:58 & $107$ \\
050416A &  &  & 0.52$\pm$0.15
& 1350$^{+2070}_{-620}$    & 0.88$\pm$0.04 & 1.05$\pm$0.08 & 11:04:45 &
$78.5$ \\
050421 & 3.05$^{+0.17}_{-0.15}$     &            & &
&            & 0.34$\pm$0.20 & 04:11:52  & $110.72$ \\
050422 &    4.97$^{+0.53}_{-0.37}$    & 272$^{+43}_{-25}$ &
0.92$^{+0.13}_{-0.16}$ &     &     & 2.15$\pm$0.94 & 07:52:42  &
$109.4$ \\
050502B &                &            & 0.8$\pm$0.2
&    ---$^d$    &            & 1.15$\pm$0.02 &
09:25:24  & $63$ \\
050505 &     &  & 0.66$^{+0.13}_{-0.12}$    & 19889$^{+5206}_{-2888}$
& 1.72$^{+0.11}_{-0.08}$ &  0.804$\pm$0.08 & 23:22:11  & $2822.2$ \\
050509A &            &            &
1.18$^{+0.21}_{-0.22}$&    See note$^a$     &             &  &
01:46:22 &  $>3000$ \\
050520 &                &            &
0.82$^{+0.48}_{-0.52}$ &            &            &
& 00:05:54  &$7661$ \\
050525A &  & &     0.98$\pm$0.05 &  641$^{+690}_{-123}$     &
1.39$^{+0.09}_{-0.04}$& 0.70$\pm$0.07    & 00:02:53  & $125.44$ \\
050603 &                &            &     &
&    1.76$^{+0.15}_-{0.07}$        & 0.71$\pm$0.10 & 06:29:01 &
$39022$\\
050607 &    2.52$\pm$0.02        &    510$\pm$50
&0.61$\pm$0.11        &  6400$\pm$900            &
1.12$\pm$0.07    & 1.15$\pm$0.11 & 09:11:22 & $99$ \\
\enddata
 \tablecomments{{\small $^a$ Only 2 points above background; $^b$ The burst time
 $T_{\rm burst}$ is given in UT. The day of the burst can be derived from the
 burst name given in the first column;
$^c$ GRB~050315 has a very complicated light curve; for a
detailed study see \cite{Vaughan05}; the spectral indices given here
correspond to the first and the third segment in the light curve,
i.e., $\beta_{\rm 1,x}$ and $\beta_{\rm 3,x}$; $^d$A flare event with
complex structure; the temporal index is for the underlying power law
decay; there is evidence for a break at $t_{\rm break,2}\sim 105\;$s,
but is is hard to determine its exact value due to flaring activity
around the same time \citep{Falcone05}. $^e$Single flare event with limited information
for the derivation of a temporal decay index \citep{Romano05}}} 
\end{deluxetable}

\clearpage 
 
\begin{deluxetable}{lcccccc}
\tablecaption{Energetics of \swift\ GRBs with known redshifts}
\tablewidth{0pt}
\tablehead{
\colhead{GRB} & \colhead{Redshift} &
\colhead{Fluence$^a$} & 
\colhead{$E_{\rm\gamma,iso}^{c*}$} &
\colhead{$E_{\rm\gamma,iso}^{c**}$} & 
\colhead{$L_{\rm X,1}^c$}  & 
\colhead{$L_{\rm X,10}^c$} \\
\colhead{} & \colhead{$z$} & \colhead{$(10^{-6}\;{\rm erg/cm^2})$} & 
\colhead{($10^{52}\;$erg)} & \colhead{($10^{52}\;$erg)} & 
\colhead{($10^{45}\;$erg/s)} & \colhead{($10^{45}\;$erg/s)}
}
\startdata
050126 & 1.29$^e$ & $1.1$ & $0.6$ & $2.2$ & $14$ & $1.2$  \\
050315 & 1.949$^f$ & $4.2$ & $5.5$ & $18$ & $780$ & $160$  \\
050318 & 1.44$^f$ & $2.1$ & $1.3$ & $3.9$ & $230$ & $6.0$  \\
050319 & 3.24$^g$ & $0.8$ & $4.0$ & $12.1$ & $550$ & $51$ \\
050401 & 2.90$^h$ & $14$ & $42$ & $137$ & $1800$  & $98$  \\
050408$^b$ & 1.236$^f$ & $2.3$ & $1.0$ & $2.9$ & $80$ & $14$  \\
050416A & 0.6535$^i$ & $0.4$ & $0.02$ & $0.09$ & $5.8$  & $0.91$  \\
050505 & 4.3$^j$ & $4.1$ & $27$ &  $89$ & $1100$ & $23$  \\
050525A & 0.606$^k$ & $20$ & $1.6$ & $3.1$ & $29$ & $1.2$  \\
050603 & 2.821$^l$ & $13$ & $31$ & $126$ &  \nodata$^d$  & $11$  \\
\enddata

\tablecomments{$^a$ Fluence is calculated between $15-350$ keV; $^b$ HETE burst, fluence is converted from $30-400$ kev using a spectral index of $\beta=-1.979$; $^c$ In all conversions we assume a cosmology with $H_0=71$ km s$^{-1}$ Mpc$^{-1}$, $\Lambda=0.27$ and $\Omega=0.73$; $^d$ XRT slewed 11 hours after trigger;$^*$ $E_{\rm\gamma,iso}$ is k-corrected and re-calculated between $100-500$ keV;$^{**}$ $E_{\rm\gamma,iso}$ is k-corrected and re-calculated between $20-2000$ keV; $^e$\citet{Berger05a}; $^f$\citet{Berger05b}; $^g$\citet{Fynbo05a}; $^h$\citet{Fynbo05b}; $^i$\citet{Cenko05}; $^j$\citet{Berger05c}; $^k$\citet{Foley05}; $^l$\citet{BerBec05}
}

\end{deluxetable}

\clearpage

\begin{deluxetable}{lcccccccccc}
\tabletypesize{\scriptsize}
\tablecaption{Energetics and Microphysical Parameters of \swift/XRT GRB Afterglows}
\tablewidth{0pt}
\tablehead{
\colhead{GRB} &
\colhead{$T_{90}$(sec)} &
\colhead{$\Delta\alpha$} &
\colhead{$\beta(p)$} &
\colhead{$k$} &
\colhead{$p$} &
\colhead{$b$} &
\colhead{$s$} & 
\colhead{$q$} & 
\colhead{$f_{\rm min}$} &
\colhead{$f_{\rm max}$} 
}
\startdata
050128 & 13.8 & 0.5 $\pm$ 0.1 & $(p-1)/2$  & 0  & 2.6 $\pm$ 0.2 & 1.4 $\pm$ 0.1 & 2.1 $\pm$ 0.2  & - 0.6 $\pm$ 0.1& 2.1 & 5.5 \\
050315 & 96.0 & 0.6 $\pm$ 0.1 & $p/2$  & 0  & 1.7 $\pm$ 0.2  & 0.9 $\pm$ 0.1  & 3.4 $\pm$ 0.1 & -0.3 $\pm$ 0.1 & 10.8  & 29.3  \\
050318 & 32.0 &  0.8 $\pm$ 0.1 & $(p-1)/2$  & 2  & 2.4 $\pm$ 0.4  & 0.9 $\pm$ 0.1  & 34.9 $\pm$ 30.9  & -0.1 $\pm$ 0.1 & 4.2  & 170.4 \\
050319 & 10.0 & 0.73 $\pm$ 0.3 & $(p-1)/2$  & 0  & 2.6 $\pm$ 0.2  & 1.4 $\pm$ 0.1  & 2.6 $\pm$ 0.5 & -0.5 $\pm$ 0.1  & 11.5   & 75.5 \\
050401 & 33.0 & 0.5 $\pm$ 0.1  & $p/2$  & 0  & 2.2 $\pm$ 0.1  & 1.0 $\pm$ 0.1 & 2.7 $\pm$ 0.1  & -0.5 $\pm$ 0.1 & 5.6 & 14.1 \\
050416a&  2.4 & 0.4 $\pm$0.1   & $p/2$  & 0  & 2.0 $\pm$ 0.1  & 1.0 $\pm$ 0.1  & 2.1 $\pm$ 0.1 & -0.6 $\pm$ 0.1  & 2.2 & 9.9   \\
050505 & 60.0 & 1.1 $\pm$ 0.1 & $(p-1)/2$  & 2  & 2.28 $\pm$ 0.2  & 0.8 $\pm$ 0.1  & -16.7 $\pm$ 4.6 & 0.3 $\pm$ 0.1 & 19.4 & 1795.0  \\
050525a&  8.8 & 0.4 $\pm$ 0.1 & $(p-1)/2$  & 2 & 2.0 $\pm$ 0.1  & 0.7 $\pm$ 0.1 & 5.9 $\pm$ 0.8  & -0.4 $\pm$ 0.1   & 2.1  & 5.9  \\
050607 & 26.5 & 0.5 $\pm$ 0.1  & $p/2$ & 0 & 2.2 $\pm$ 0.2 & 1.1 $\pm$ 0.1 & 2.5 $\pm$ 0.1 & -0.5 $\pm$ 0.1 & 3.4 & 14.1\\
\enddata
\end{deluxetable}

\end{document}